\documentclass{aa}

\usepackage{graphicx}
\usepackage{txfonts}
\usepackage{amsmath}            
\usepackage{placeins}           
\usepackage{xcolor}
\usepackage{bm}
\usepackage{tikz}               
\usetikzlibrary{shapes.geometric, arrows.meta, positioning}

\begin{document}

   \title{Gaia as a $0\farcs1$ point-source survey across the sky:}

   \subtitle{dynamical scene reconstruction from one‑dimensional scans}

   \titlerunning{Gaia as a 0.1 arcsec point-source survey across the sky}

   \author{Sahar Shahaf\inst{1}
   \and {David W. Hogg}\inst{1,2,3}
   \and {Kareem El-Badry}\inst{1,4}
   \and {Hans-Walter Rix}\inst{1}
    \and {Ren\'e Andrae}\inst{1}
        }

   \institute{Max-Planck-Institut f\"ur Astronomie (MPIA), K\"onigstuhl 17, 69117 Heidelberg, Germany.
    \and
    Department of Physics, New York University, 726 Broadway, New York, NY 10012, USA.
    \and
    Center for Computational Astrophysics, Flatiron Institute, 162 Fifth Ave., New York, NY 10010, USA.
    \and
    Department of Astronomy, California Institute of Technology, 1200 E. California Blvd, Pasadena, CA, 91125, USA.
        }

   \date{Received ---; accepted ---}

  \abstract{
Gaia has observed every detected source on the sky 50 to 100 times, each time through a window of about 1 by 2 arcseconds at a different scan angle. The resulting one-dimensional flux profiles to be released in Gaia DR4 encode the two-dimensional arrangement of nearby point sources at ${\sim}$0.1 arcsecond resolution. Repeated observations also constrain their relative motion. We present a method to model the scene with a mixture of moving point sources. Complementing Gaia DPAC's source-environment reconstruction, our approach focuses on close, blended companions and explicitly models their relative motion across the observing baseline. We first fit each window with a few point sources and obtain their along-scan positions. We then triangulate these positions from different scan angles to identify candidate sources on the sky. A joint forward model of all windows measures positions, fluxes, relative proper motions, and parallax differences, including when the components are blended.
In simulations, we recover companions at separations smaller than the instrumental profile width. For a moving companion at ${\sim}\,75$ mas separation with a flux ratio of ${\sim}\,0.2$, we obtain a separation precision of ${\sim}\,0.1$ mas and a relative proper-motion precision of ${\sim}\,0.1$ mas\,yr$^{-1}$ per component, even in a scene ``contaminated" by a third source. These precisions assume a known instrumental profile and independent Gaussian noise. We also recover differential parallax for a chance-aligned source. Modeling nearby contaminants prevents significant biases. With a Gaia-like instrumental profile, sampling and scan law, we show a sensitivity of ${\sim}\,5$ mag in contrast at separations of $50$ to $150$ mas, and we discuss its dependence on the color term of Gaia's line-spread function. The method could extend the census of close companions and provide relative astrometry, and so dynamical information, along short orbital arcs. External measurements could also guide retrospective searches in Gaia's windows.
}
\maketitle

\nolinenumbers
\section{Introduction}
\label{sec:intro}

High-angular-resolution imaging separates close sources and reveals faint companions hidden in the light of brighter neighbors. Achieving this resolution over large areas of sky is costly, so most observations cover selected targets or limited fields.

Wide-field ground-based optical surveys generally operate near the atmospheric seeing limit. The Pan-STARRS1 survey covers roughly $30{,}000$ square degrees with typical image quality of about $1.2$ arcseconds \citep{Chambers2016}, while the Rubin Observatory is designed to survey 18,000 square degrees with a median delivered image quality near 0.7 arcseconds \citep{Ivezic2019}. Space telescopes are unaffected by seeing conditions. Euclid reaches about 0.18 arcseconds over 14,000 square degrees, although its wide survey largely avoids the Galactic plane \citep{EuclidCollaboration2025}. Its nominal exposures also saturate on point sources brighter than $m_\mathrm{AB}\approx18$ \citep{EuclidCollaboration2025}. At finer resolution, sky coverage becomes more limited. HST and JWST provide resolutions of order 0.05 to 0.1 arcseconds, but only in small fields. Adaptive optics reaches tens of milliarcseconds for suitable targets, and long-baseline interferometry reaches milliarcsecond scales for a smaller sample still.

Close stellar companions are common and span a wide range of separations, but most are unresolved in low-resolution images. Detecting them requires techniques whose sensitivity depends on separation and contrast \citep{Moe2017}. Repeated high-resolution observations can both identify companions and measure their relative motion \citep{Vrijmoet2026}. These measurements provide dynamical information even when the observing baseline covers only a small fraction of an orbit \citep{Pearce2015}. A survey with a known selection function would constrain companion occurrence as a function of separation and flux ratio, in the field as in clusters and associations.

Gaia offers repeated observations of every source on the sky at high angular resolution along each scan, over a brightness range from $G\approx3$ to $G\approx20.7$ \citep{GaiaCollaboration2016}. Its rectangular aperture produces an anisotropic diffraction pattern, narrowest in the along-scan direction. The corresponding line-spread function (LSF) has a characteristic FWHM of ${\sim}\,100$ mas \citep{Rowell2021} and is sampled at $59$ mas per pixel \citep{GaiaCollaboration2016}. Its shape depends on source color, position in the focal plane, and time \citep{Rowell2021}. For sources fainter than $G\simeq13$, the astrometric windows are normally binned over $12$ across-scan pixels and transmitted as one-dimensional profiles \citep{Rowell2021}. Each profile constrains source positions along one direction. Observations at different scan angles together deliver information about their two-dimensional arrangement.

Reconstructing source environments from scanning astrometry has precedents in Hipparcos \citep{Quist1999} and Gaia \citep{Harrison2011,Harrison2023}. Within Gaia DPAC, the Source Environment Analysis Pipeline \citep[SEAPipe;][]{Harrison2023} relies on a weighted stack of the window samples in the primary source's co-moving frame. The weighting prefers low flux levels, so samples contaminated by the primary's flux are down-weighted and the stacked image keeps the sky level around the primary \citep{Harrison2011}. Sources identified in the image are fitted to the window samples. In the image-subtraction version of the pipeline, they are then removed and the search is repeated. The sources that SEAPipe detects around main-catalogue sources are planned for release in the Gaia DR4 \texttt{gaia\_source\_environment} table.\footnote{See \url{https://www.cosmos.esa.int/web/gaia/dr4} for the expected release contents, subject to processing and validation.}

Here we use Gaia's one-dimensional windows to identify close, blended companions and measure their relative motion. SEAPipe and our approach differ in how they find candidates. The SEAPipe stack needs no model of the instrumental profile, so it is robust to its variations, and it essentially aggregates flux. The instrument model and the motions enter only when the sources identified in the stack are fitted to the window samples. We instead model the LSF already in the reduction of every window and accept the costs of that model, discussed in Section~\ref{sec:precision}, to achieve improved localization and to monitor the scene in time. Beyond the candidate search, this paper recovers companions at separations below the FWHM of the LSF, and calibrates the detection limits and the color bias for a Gaia-like LSF. 

For a scene of a few moving point sources, we fit each window separately and obtain along-scan detections. Each detection defines a line of possible positions on the sky. Accumulating these lines across scan angles identifies candidate sources at their intersections. We refer to this mapping from detections to sky lines as backprojection. This geometry is closely related to the Radon transform used in tomography \citep{Radon1917,Kak2001}, but here we accumulate fitted source detections to identify candidate point sources. We account for relative motion by propagating detections to a common reference epoch using a fitted motion model. A joint forward model of the original windows then measures the components' positions, fluxes and motions.


\section{The transit measurement}
\label{sec:measurement}
We begin with a single one-dimensional Gaia window, which records the local stellar scene along the instantaneous scan direction. The observed profile is blurred by the instrument's LSF \citep{Rowell2021}. Throughout this section, the along-scan coordinate, $\eta$, is measured in milliarcseconds relative to a reference point of the window. Its mapping onto the sky is developed in Section~\ref{sec:backprojection}. The choice of this point is a convention. It may be the reference coordinate supplied with the Gaia window or the fitted position of the primary, and the method requires only that it is known for each window. For simplicity, the first examples place it at the primary. In the more elaborate treatment of Section~\ref{sec:crowded}, the primary's position is not fixed and is fitted together with the other sources.

\subsection{Point sources along a single scan}
\label{sec:meas-single}
Consider a bright primary of flux $f_1$ at along-scan position $\eta_1$, together with a single faint contaminant of flux $f_2$ at $\eta_2$. We denote the secondary-to-primary flux ratio by $\mathcal{S}=f_2/f_1$. The window contains the contribution of both components,
\begin{equation}
s = f_1\,\mathcal{P}\left(\eta-\eta_1\right)
+ f_2\,\mathcal{P}\left(\eta-\eta_2\right)
+ \text{noise},
\label{eq:forward-two}
\end{equation}
where $\mathcal{P}$ is the along-scan line-spread function, normalized such that a well-contained source sums to unity over the sampled grid. Each source enters equation~\eqref{eq:forward-two} through its LSF, the one-dimensional image of a point source marginalized over the across-scan direction \citep{Rowell2021}. All modeled sources lie within the across-scan extent of the window. Brighter sources are observed in two-dimensional windows, which we discuss in Section~\ref{sec:discussion}.

When the projected separation of the two stars along the scan is large compared with the width of the LSF, their profile is double-peaked \citep{Penoyre2026}, and the position of each component can be measured. As the projected separation decreases, the two peaks merge into a single blended profile.
A blended profile can be summarized by its photocenter. For a fully blended pair, the along-scan position is
\begin{equation}
\eta_\mathrm{pc}
= \frac{f_1\eta_1+f_2\eta_2}{f_1+f_2}
= \eta_1 + \frac{f_2}{f_1+f_2}\left(\eta_2-\eta_1\right).
\label{eq:photocentre-pull}
\end{equation}
The photocenter therefore lies between the two sources, displaced from the primary by a fraction of their separation set by the flux ratio \citep[e.g.,][]{Penoyre2020, Shahaf2019}. It is not generally the same as either the peak of the blended profile or the position obtained by fitting a single-source model.

As Gaia's scan direction changes, the projected separation changes, altering the blended profile and the position inferred from a single-source fit. Such scan-angle-dependent biases can produce excess astrometric residuals and spurious periodic signals \citep{Penoyre2020, Holl2023, ElBadry2024}. Here, we model the contributing sources explicitly, using the changing projections to constrain their positions and fluxes.
We model the window as a superposition of $K$ point sources,
\begin{equation}
s(\eta)=\sum_{k=1}^{K}f_k\,\mathcal{P}(\eta-\eta_k)+\varepsilon(\eta).
\label{eq:forward}
\end{equation}
Stacking the samples into a vector gives
\begin{equation}
\vec{s}=\mathbf{A}(\boldsymbol{\eta})\,\vec{f}+\vec{\varepsilon},
\label{eq:forward-matrix}
\end{equation}
where $\vec{f}=(f_1,\dots,f_K)$ contains the source fluxes, $\vec{\varepsilon}$ the noise, and each column of $\mathbf{A}$ the sampled profile of a source at one trial position. For the pair of equation~\eqref{eq:forward-two}, $\vec{f}$ contains two entries, representing the flux of the primary and secondary. At fixed positions, the model is linear in the non-negative source fluxes. We assume that the background has been subtracted. Application to released Gaia windows may require fitting a residual background term as well. Recovering the source positions and fluxes from each window is the goal of the following procedure.

\subsection{Detecting sources in a single window}
\label{sec:meas-nnls}
Modeling the along-scan profile requires the position of each source and the shape of its LSF. Both enter the model nonlinearly, so we begin with an approximate, linear step. We construct $\mathbf{A}$ as a dense dictionary of LSF columns, placed at the nodes of a grid finer than the pixel sampling. The LSF shape follows from the color of the source \citep{Rowell2021}. Adding shapes for nearby colors at each node allows for an uncertain color and for companions whose color differs from the primary's. With the positions and shapes fixed, equation~\eqref{eq:forward-matrix} is linear in the fluxes and is solved directly in each window. The resulting flux distribution shows where sources are likely to lie and provides the starting point for the fits described below. In the examples of this paper, each node carries a single shape.

We assume independent Gaussian sample noise with known fixed variances. The normalization of the likelihood then does not depend on the fitted parameters, so maximizing it is equivalent to minimizing $\chi^2$. We write out the $\chi^2$ terms to show how overlapping LSF columns interact, and this interaction motivates the non-negativity constraint on the fluxes that regularizes the fit. To simplify the notation, we take the variances to be equal, and unequal known variances only require dividing each sample and its model prediction by the corresponding noise standard deviation. Then $\chi^2$ is proportional to the squared norm of the residuals,
\begin{equation}
    \chi^2 \propto  \lVert\vec{s}-\vec{m}\rVert^2= \underbrace{\vec{s}^{\intercal}\vec{s}}_{\text{data}} - 2\underbrace{\vec{m}^\intercal\vec{s}}_{\text{overlap}} + \underbrace{\vec{m}^\intercal  \vec{m}}_{\text{model}},
\end{equation}
where $\vec{m}\equiv\mathbf{A}\vec{f}$ is the parameterized model vector. The data norm $\vec{s}^{\intercal}\vec{s}$ is constant with respect to the fitted fluxes.

Away from the window boundaries, the shifted LSF columns have approximately the same squared norm. The model norm can therefore be written as
\begin{equation}
\vec{m}^{\intercal}\vec{m}
\simeq
a\,\lVert\vec{f}\rVert^2
+
2\sum_{k<\ell}a_{k\ell}f_{k}f_{\ell},
\end{equation}
where $\vec{a}_k$ is column $k$ of $\mathbf{A}$, $a$ is the approximate squared norm of an LSF column, and $a_{k\ell}\equiv\vec{a}_{k}^{\intercal}\vec{a}_{\ell}$ measures the overlap between columns $k$ and $\ell$. Since the LSF is non-negative, $a_{k\ell}\geqslant0$, with the largest overlaps occurring between nearby candidate positions. Allowing coefficients of opposite sign would let these columns partially cancel one another.

Physical source fluxes are non-negative. Enforcing this constraint removes such cancellations and has a regularizing effect. It restricts the allowed solutions and can favor fits with fewer active components. We therefore solve the non-negative least-squares problem
\begin{equation}
\hat{\vec{f}}
=
\arg\min_{\vec{f}\geqslant0}
\chi^2(\vec{f}\,;\,\mathbf{A},\mathbf{\Sigma}).
\label{eq:nnls}
\end{equation}
Here $\mathbf{\Sigma}$ is the noise covariance matrix of the window samples, diagonal for independent noise. Additional regularization, such as an $\ell_1$ penalty on the fluxes, could further constrain the dense dictionary, but is not used here. We return to sparse recovery in Section~\ref{sec:discussion}. Non-negativity alone does not determine the number of sources, because adding candidate components can still reduce the fitted residual.

The solution of equation~\eqref{eq:nnls} is the approximate, linear step of the first paragraph, with the positions restricted to the nodes of the grid and a fixed LSF shape at each node. The grid is dense, but a source generally lies between nodes, so its flux is spread over the neighboring nodes rather than placed at its true position. We therefore use this solution only to initialize the continuous source fits. Nodes with nonzero flux mark candidate regions, and neighboring nodes separated by no more than one pixel are grouped into clusters. Neither the number of such nodes nor the number of clusters is a source count, because one source may populate several neighboring nodes, and blended sources may occupy the same cluster.

Starting from these nodes and clusters, we fit models with increasing numbers of sources, in which the positions vary continuously and the non-negative fluxes are solved at each step. An additional source is accepted when the decrease in $\chi^2$ it produces, $\Delta\chi^2$, exceeds a simulation-calibrated threshold $\Lambda_{\rm w}$. Appendix~\ref{app:reduction} gives the fitting and calibration procedures.
Each scan window is reduced to a \emph{transit detection set},
\begin{equation}
\label{eq:detection_set}
D_{j}
=
\left\{
(\hat{\eta}_k,\hat{f}_k,\sigma_{\eta,k})
\,\middle|\,
k=1,\ldots,K_{j}
\right\}.
\end{equation}
Here $j$ labels the window, observed at time $t_j$ and scan angle $\psi_j$, $K_{j}$ is the number of components supported by the per-window fit, and $\sigma_{\eta,k}$ is the formal along-scan uncertainty obtained from the curvature of the profiled likelihood at the fitted solution. The detection set includes all fitted components.

\begin{figure}
\centering
\includegraphics[width=0.95\columnwidth]{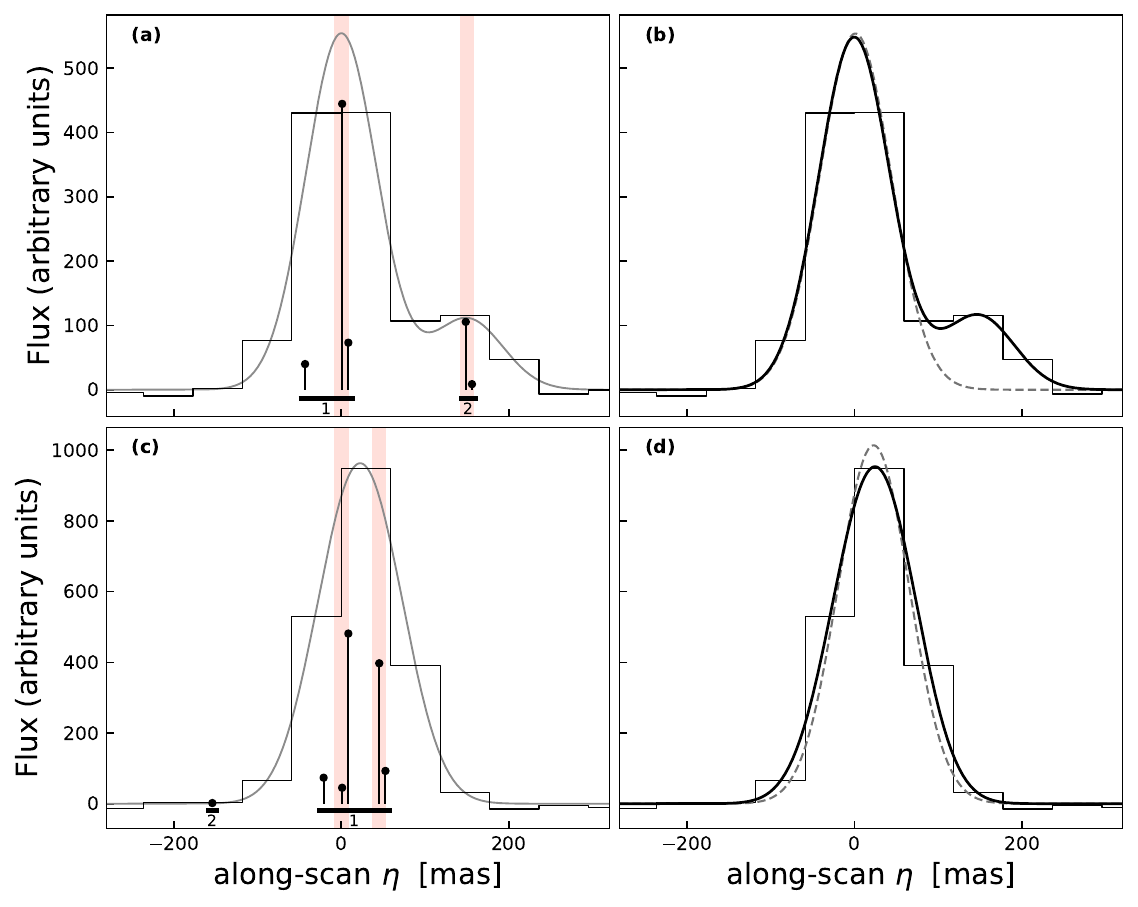}
\caption{Per-window reduction for a resolved pair separated by $150$ mas with a flux ratio of $0.2$ (top) and an equal-flux pair separated by $45$ mas (bottom). \emph{Left, panels (a) and (c):} the stepped profile shows the measured samples, the gray curve shows the noise-free profile, and the shaded vertical bands mark the injected source positions. Black stems show the non-negative fluxes fitted on the position grid. Numbered horizontal bars identify clusters of nodes with nonzero flux. These clusters provide starting positions for the continuous fits, not a source count. \emph{Right, panels (b) and (d):} the fitted two-source profile (solid black) is compared with the best single-source profile (dashed gray). In the close blend, the observed profile has only one peak, but its shape supports a two-source fit.}
\label{fig:per_window_flow}
\end{figure}

Figure~\ref{fig:per_window_flow} illustrates the reduction for a resolved companion and a close blend. In the close blend, the profile has a single peak, yet its shape supports a two-source fit. Figure~\ref{fig:scan_to_skyline}a shows how the projected separation, and with it the blend, changes with scan angle for a simulated pair.

\section{Triangulation maps}
\label{sec:backprojection}

The reduction in Section~\ref{sec:measurement} provides a detection set $D_{j}$ for each observation at time $t_j$ and scan angle $\psi_j$. Each window tells us how many components it supports, but not how many sources the scene contains. A pair resolved at one scan angle can be blended at another, and a window can also yield a spurious detection. Deciding how many sources are present, and where they are, is the main optimization challenge of the reconstruction. The number of sources is a discrete choice, so it cannot be found by following derivatives. Our aim is a joint fit to all windows, in which models with one, two, three, or more sources are compared and an additional source is kept only when it improves the fit by more than a calibrated threshold. Such a fit, however, finds the right configuration only if it starts near it, and the following sections build that starting point.

We therefore add an intermediate step before fitting the windows. Each detection places its source somewhere along a line on the sky, perpendicular to the scan, and lines from different scans that belong to the same source cross at its position. We accumulate these lines in a triangulation map, whose peaks propose candidate sources and the detections that belong to each. Each candidate seeds the joint fit to the windows, which decides whether the data support it. The map pools the positions measured in single windows, which are the quantities the reconstruction needs, and gathers the detections of each star from all scans. Unlike an image, it adds up positions rather than flux.

\subsection{From scan angles to sky positions}
\label{sec:bp-geometry}
We work in a frame following the primary's predicted astrometric track, with along-scan positions measured relative to that track. For a bound companion, this approximately removes the parallax and systemic proper motion shared by the pair, leaving relative orbital motion.

We first consider a scene whose relative positions and component fluxes remain constant over the observing baseline. The companion's sky offset is
\begin{equation*}
(\Delta\alpha^{*},\Delta\delta)=\rho\,(\sin\varphi,\cos\varphi),
\end{equation*}
where $\Delta\alpha^{*}\equiv\Delta\alpha\cos\delta$, $\rho$ is the angular separation, and $\varphi$ is the position angle measured from north through east. A scan at angle $\psi$, measured in the same convention, places the companion at the along-scan coordinate
\begin{equation}
\eta(\psi)
=\Delta\alpha^{*}\sin\psi+\Delta\delta\cos\psi
=\rho\cos(\psi-\varphi)\equiv\rho_\parallel.
\label{eq:projection}
\end{equation}
We call a plot of along-scan position $\eta$ against scan angle $\psi$ a \emph{sinogram}. A source at a fixed sky position traces a sinusoid in this plot. The projected separation $\rho_\parallel$ is largest when the scan is aligned with the pair and vanishes when it is perpendicular. The same pair can therefore appear well separated in some windows and strongly blended in others, even when its sky position remains unchanged.

A detection at along-scan coordinate $\hat{\eta}$ in transit $j$ is consistent with every sky offset that projects onto the scan axis at $\hat{\eta}$. These offsets form a line on the sky,
\begin{equation}
\Delta\alpha^{*}\sin\psi_j+\Delta\delta\cos\psi_j=\hat{\eta},
\label{eq:line}
\end{equation}
which is perpendicular to the scan direction, with a transverse uncertainty set by the along-scan measurement error. Two detections of the same source at non-parallel scan angles constrain a two-dimensional position. Additional observations improve the precision and help distinguish a consistent source position from chance intersections. The positional precision depends on both the measurement uncertainties and the angular coverage.

\begin{figure*}
\centering
\includegraphics[width=\textwidth]{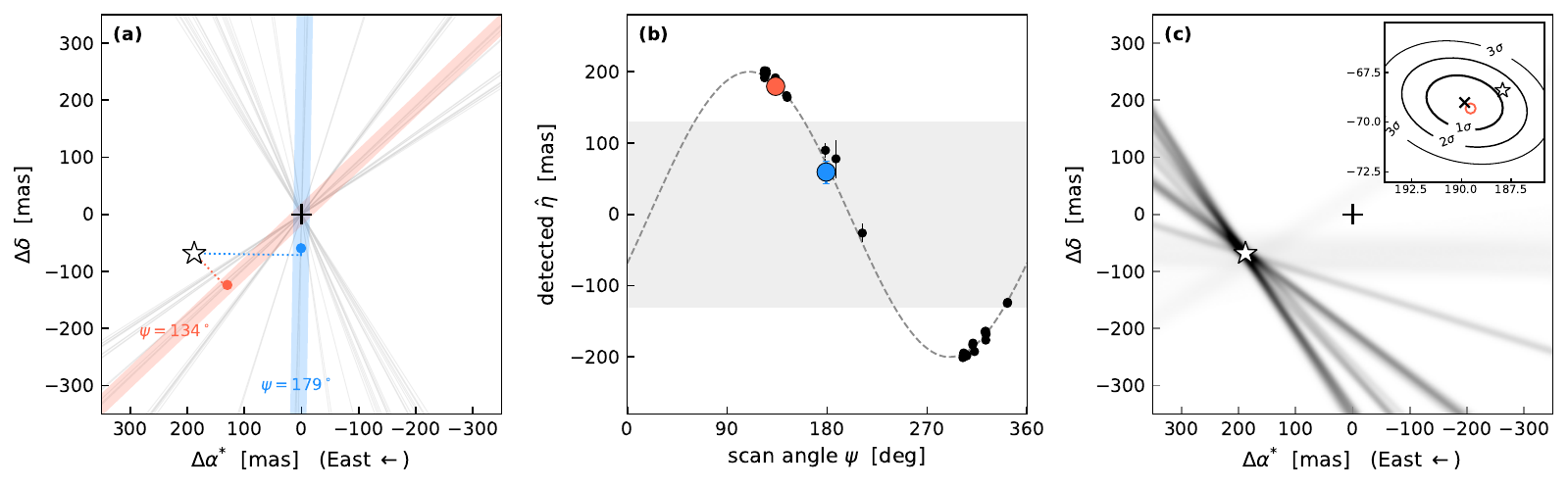}
\caption{From along-scan detections to a two-dimensional source position. \emph{(a)} Scan geometry for the simulated pair, with the primary at the origin (cross) and the companion marked by the star. Each line through the origin is the scan axis of one transit, and two transits are highlighted. The dotted segment drops the companion perpendicularly onto the scan axis, and the dot marks the along-scan position measured from the window. Figure~\ref{fig:per_window_flow} shows how a single window is reduced to such along-scan positions. \emph{(b)} Along-scan detections as a function of scan angle, with formal positional uncertainties. The dashed curve is the injected track, $\rho\cos(\psi-\varphi)$. The gray band marks projected separations for which the noise-free blended profile of this pair, with a flux ratio of $0.2$, has a single peak. This is not a boundary for detecting a second source. \emph{(c)} The triangulation map combines bands around the sky lines allowed by the detections, with widths set by their positional uncertainties. The common intersection identifies a candidate companion position. The inset shows the subsequent joint window fit described in Section~\ref{sec:bp-measure}, with profiled $\Delta\chi^2$ contours corresponding to $1$, $2$, and $3\sigma$. The circle marks the map-associated starting position, the cross the maximum-likelihood position, and the star the injected position.}
\label{fig:scan_to_skyline}
\end{figure*}

\subsection{Fitting the sinogram}
\label{sec:sinogram-static}
When the scene contains a single companion, every detection not assigned to the primary belongs to it, and its sky position can be fitted directly from the sinogram. Under the static-scene assumption and Gaussian positional errors,
\begin{equation}
\hat{\eta}_i=a_1\sin\psi_i+a_2\cos\psi_i+\varepsilon_i,
\qquad
\varepsilon_i\sim\mathcal{N}(0,\sigma_{\eta,i}^{2}).
\label{eq:sinogram-fit}
\end{equation}
Comparison with equation~\eqref{eq:projection} identifies $a_1$ and $a_2$ with the companion's sky offsets $\Delta\alpha^{*}$ and $\Delta\delta$. They are estimated by weighted least squares, using the uncertainties of the detections. This fit is simple and linear, but it is biased by blending, as we discuss in Section~\ref{sec:sinogram}. Figure~\ref{fig:scan_to_skyline}a illustrates the scan geometry, while panel~(b) shows the corresponding along-scan detections as a function of scan angle.

The angular separation and position angle derived from the sinogram are
\begin{equation}
\hat{\rho}=\sqrt{a_1^2+a_2^2},
\qquad
\hat{\varphi}=\operatorname{atan2}(a_1,a_2),
\label{eq:polar-elements}
\end{equation}
where the two-argument arctangent selects the quadrant from the signs of the fitted offsets. Uncertainties follow by propagating the covariance of $a_1$ and $a_2$. This linear error propagation is appropriate when the positional uncertainty is small compared with the separation. Near the origin this approximation fails. Because $\hat{\rho}$ cannot be negative, noise in $a_1$ and $a_2$ raises its mean above the true separation, and the position angle becomes poorly constrained.

\subsection{Triangulation}
\label{sec:bp-detect}
The sinogram fit uses the detections in scan coordinates, but the same detections can also be represented on the sky. Each detection contributes along the line defined by equation~\eqref{eq:line}, and contributions from the same source accumulate near a common intersection, as illustrated in Figure~\ref{fig:scan_to_skyline}c. We call this representation the \textit{triangulation map}. Detections assigned to the primary are omitted from the map, and the fitted fluxes are not used to weight the contributions.

The map measures how strongly each sky position is supported by the detections, with more weight for more precise ones. We evaluate it on a milliarcsecond sky grid. Each detection contributes a band around its sky line, with a Gaussian profile across the band whose width is the detection's along-scan uncertainty. For a sky position $\vec{x}$ at a perpendicular distance $\delta_{\vec{x},i}$ from the line of detection $i$, the map value is
\begin{equation}
M_{\vec{x}}
=\sum_i
\frac{1}{\sqrt{2\pi}\sigma_{\eta,i}}
\exp\left(-\frac{\delta_{\vec{x},i}^{2}}{2\sigma_{\eta,i}^{2}}\right).
\label{eq:hough-map}
\end{equation}
Every band carries the same total weight, so less precise detections contribute broader bands with lower peaks. Adding these contributions highlights positions supported by several detections, without first assigning them to individual sources.

The map is needed when the scene contains several sources. Their number is then not known, and windows can also yield spurious detections, so the detections cannot be assigned to sources in advance. Additional sources and spurious detections add lines that do not share the companion's intersection. The strongest peak away from the primary proposes a candidate, and the detections whose lines pass through it within their uncertainties are associated with it. The map is used to locate candidates and is not a likelihood surface.

The triangulation map is also not an intensity image. Radon inversion reconstructs a brightness distribution from its projections, while SEAPipe identifies candidate sources in a reconstructed image \citep{Kak2001,Harrison2023}. The peaks of our map instead represent agreement between positions measured in individual windows. Appendix~\ref{app:radon} describes the connection to the Radon transform and compares the map with Radon inversion and with faststack (Figures~\ref{fig:radon} and~\ref{fig:pictures}).

\section{Scene reconstruction}
\label{sec:bp-measure}
We consider a scene containing a primary and one companion. We first extend the projection geometry to include relative motion and fit it to the detections, then describe the joint forward fit used to measure the component positions, fluxes, and motions from the original windows. Scenes containing additional sources are considered in Section~\ref{sec:crowded}.

\subsection{A dynamical scene}
\label{sec:dynamics}

So far, we have assumed that the scene is static in the primary's co-moving frame. In a bound system, the companion's offset from the primary changes with time, so its detections depart from a single sinusoid in the sinogram, and its sky lines no longer meet at a common point in the triangulation map.

\begin{figure*}
\centering
\includegraphics[width=\textwidth]{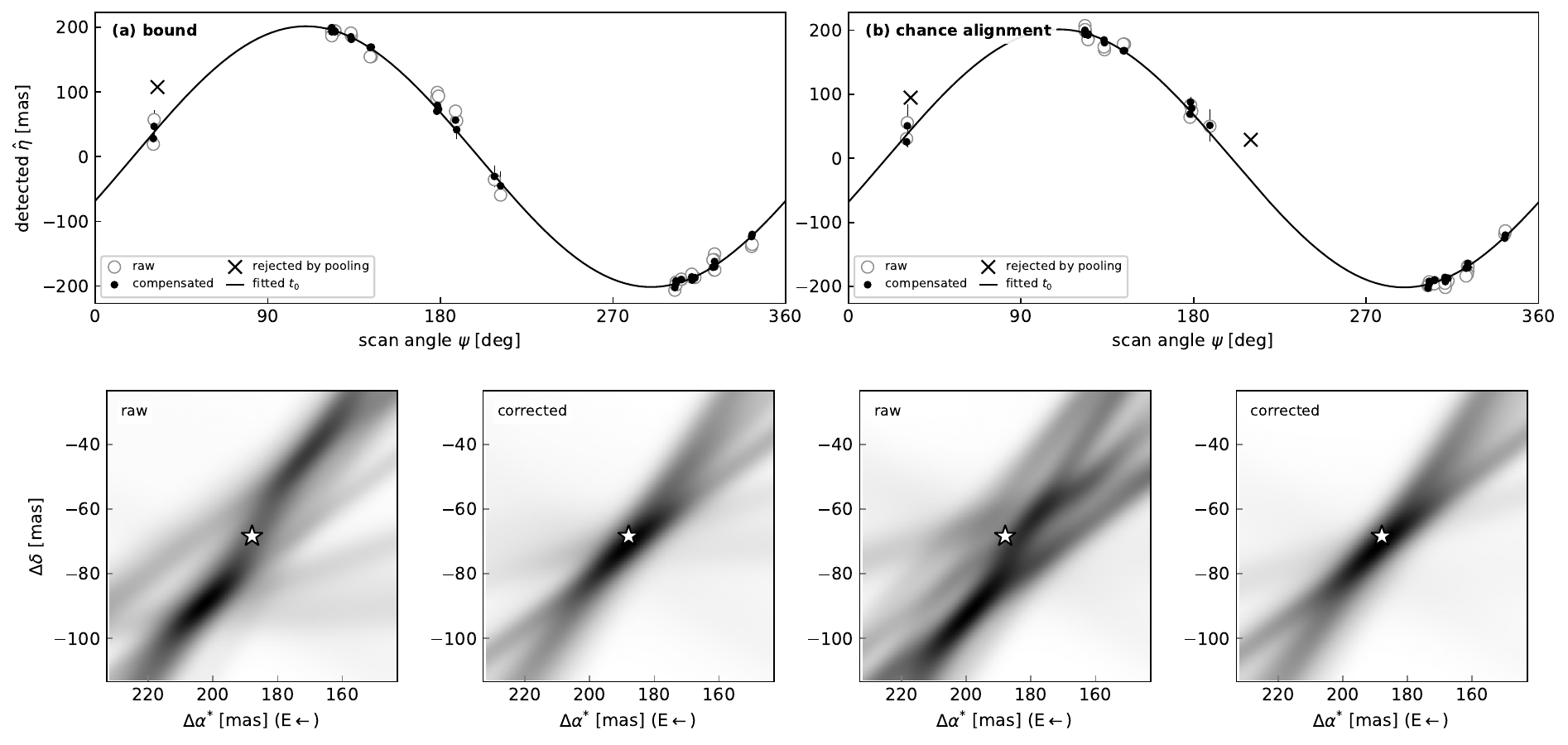}
\caption{Motion compensation for a bound companion (a) and a chance-aligned background source (b). Both scenes have a separation of $200$ mas at the reference epoch, a flux ratio of $0.3$, and the uniform noise of Section~\ref{sec:static-example}, sampled at the same $47$ GOST-predicted transits as Figure~\ref{fig:scan_to_skyline}. \emph{Top:} along-scan detections before compensation (open circles) and after removing the fitted relative motion and, for the background source, differential parallax (filled points). Crosses mark detections rejected by pooling. The solid curve shows the fitted scene at the reference epoch. The compensation uses the parameters from the joint window fit. \emph{Bottom:} triangulation maps before and after compensation for each scene. Stars mark the injected position at the reference epoch. Accounting for motion brings the projected detections into closer agreement and concentrates the map near that position.}
\label{fig:dynamics}
\end{figure*}

Linear relative motion keeps the fit to the detections linear. About a reference epoch $t_0$, write the companion's offset from the primary as
\begin{equation}
\Delta\alpha^{*}(t) = \alpha_0^{*} + \mu_{\alpha^*}\,\tau \quad\text{and}\quad
\Delta\delta(t)   = \delta_0 + \mu_{\delta}\,\tau,
\label{eq:linear-motion}
\end{equation}
where $\tau\equiv t-t_0$.

\subsection{Fitting the sinogram of a moving source}
\label{sec:sinogram}
Projecting the motion of equation~\eqref{eq:linear-motion} onto the scan direction using equation~\eqref{eq:projection}, the along-scan position satisfies
\begin{equation}
\eta(\psi,t) = \alpha_0\sin\psi + \delta_0\cos\psi
           + \mu_{\alpha^*}\,(\tau\sin\psi)
           + \mu_{\delta}\,(\tau\cos\psi).
\label{eq:sinogram-dynamics}
\end{equation}
This expression is still linear in the four unknowns $(\alpha_0,\,\delta_0,\,\mu_{\alpha^*},\,\mu_{\delta})$, and the static design of equation~\eqref{eq:sinogram-fit} simply gains two additional columns, $\tau\sin\psi$ and $\tau\cos\psi$. A single weighted least-squares solve over the detection set returns the position at $t_0$ and the relative proper motion, with no nonlinear search. The static fit is the $\mu=0$ special case. A constant relative acceleration adds two further linear columns, whereas strongly curved arcs require a full Keplerian model and a nonlinear fit.

The linear approximation assumes that the orbital arc is nearly straight over the mission baseline. A chance-aligned source may have a parallax different from that of the primary, introducing an additional term $\Delta\varpi\,f_{\varpi}$,
\begin{equation}
\begin{aligned}
\eta(\psi,t, f_{\varpi})  &= \alpha_0\sin\psi + \delta_0\cos\psi\\
                &+ \mu_{\alpha^*}\,(\tau\sin\psi) + \mu_{\delta}\,(\tau\cos\psi)  + \Delta \varpi f_{\varpi},
\label{eq:sinogram-dynamics-extended}
\end{aligned}
\end{equation}
where $f_{\varpi}$ is the along-scan parallax factor. The relative proper motion is the dynamical quantity we fit in the following sections.

Fitting the sinogram is simple and inexpensive, but it is biased. Windows in which the companion was not separated contribute no detection, so the detections favor the scan angles at which the pair is well separated. Near the blending limit, whether a second component is accepted depends on the noise realization, which biases the retained positions, and the uncertainties of the detections do not include this bias. Section~\ref{sec:moving-example} quantifies this bias for a moving pair.

When several sources are present, the detections are associated with them through the triangulation map of Section~\ref{sec:bp-detect}, which for a moving source must account for its motion. The fitted motion can be removed, propagating each detection back to the reference epoch,
\begin{equation}
\label{eq: eta_dyn}
\hat{\eta}^{\mathrm{dyn}}_i=
\hat{\eta}_i-(
\mu_{\alpha^*}\tau_i\sin\psi_i +
\mu_{\delta}\tau_i\cos\psi_i + \Delta \varpi f_{\varpi}),
\end{equation}
so that the corrected detections follow the track expected at the reference epoch $t_0$. More detailed motion models can be treated in the same way. After the correction, the allowed sky lines intersect again near the companion's position at $t_0$. We call this construction of a triangulation map from motion-corrected detections \emph{dynamical backprojection}. Figure~\ref{fig:dynamics} shows the triangulation maps before and after this correction for a bound companion and a chance-aligned source (Section~\ref{sec:moving-example}).

Either the fit to the detections or the peak of the triangulation map can initialize the joint fit to the windows, which removes the bias described above. For the pairs of Sections~\ref{sec:static-example} and~\ref{sec:moving-example}, we start from the fit to the associated detections, and for the three-source scene of Section~\ref{sec:crowded}, from the map peak.

\subsection{Forward modeling}
The joint fit models the original windows, including those in which the companion was not detected, and provides the parameter estimates and uncertainties that we report.

The parameter vector $\Theta$ contains the fitted positions and, when included, motion parameters of the two sources relative to the adopted reference track. For each trial $\Theta$, we calculate their along-scan positions $\eta_{1,j}$ and $\eta_{2,j}$ at the scan angle and epoch of window $j$. These positions determine the two shifted LSF columns of $\mathbf{A}_j$ through equation~\eqref{eq:forward-matrix}. The component fluxes are collected in a non-negative vector $\vec f=(f_1,f_2)^\intercal$, shared by all windows under the assumption that both sources have constant flux.
For independent windows with Gaussian noise, the joint log-likelihood is, up to an additive constant,
\begin{equation}
\ln\mathcal{L}(\Theta,\vec f)
= -\frac{1}{2}\sum_j
\left\lVert\mathbf{\Sigma}_j^{-1/2}
\left(\mathbf{s}_j-\mathbf{A}_j(\Theta)\vec f\right)\right\rVert^2,
\label{eq:joint-likelihood}
\end{equation}
where $\mathbf{s}_j$ contains the measured samples and $\mathbf{\Sigma}_j$ is their noise covariance matrix. At each trial $\Theta$, the two shared fluxes are fitted by non-negative least squares over all windows. Profiling out the fluxes leaves a nonlinear search over the fitted position and motion parameters, initialized from the triangulation map and the associated along-scan detections. The treatment of the primary's reference-track uncertainty is specified in each example.

External measurements of component positions or motions can be incorporated through a prior $\pi(\Theta)$. The joint fit then maximizes the log-posterior,
\begin{equation}
\ln {\rm Pr}\left(\Theta,\vec f\mid\{\mathbf{s}_j\}\right)
=\ln\mathcal{L}(\Theta,\vec f)+\ln\pi(\Theta).
\label{eq:joint-posterior}
\end{equation}
Since the prior does not involve the fluxes, they remain profiled out by non-negative least squares. Flux constraints can also be included, but would require adapting the profiled flux fit used here to optimize the likelihood and flux prior together.

\subsection{Example: a static pair}
\label{sec:static-example}
We demonstrate the static reconstruction with a primary and companion separated by $200$ mas at a position angle of $110^\circ$, with a flux ratio of $0.2$. We use $47$ transits predicted for the center of the Pleiades during the DR4 time span by ESA's Gaia Observation Forecast Tool (GOST).\footnote{\url{https://gaia.esac.esa.int/gost/}} The windows are simulated in one dimension. For each transit, the sky offsets of the sources are projected onto the scan axis with equation~\eqref{eq:projection}, and the window samples are computed from the same LSF model that the fits use, a Gaussian of FWHM $100$ mas sampled at $59$ mas over $18$ samples. All simulations in this paper use windows of $18$ samples. The windows carry independent Gaussian noise with the same standard deviation in every sample, set so that the primary has a signal-to-noise ratio of $100$ in each window, defined as its total flux divided by the noise per sample. We refer to this as the uniform noise. It is a toy model that shows the geometry of the method, and Gaia's noise and LSF enter in Sections~\ref{sec:crowded-example} and~\ref{sec:lsf-errors}--\ref{sec:detection}. No two-dimensional image is simulated. 

For this example, the primary's reference track is treated as known and its position is fixed at the origin. This is an idealization. A reference track from Gaia's single-source astrometry would be pulled toward the companion by an amount that changes with scan angle. The pull is small where the pair is resolved and approaches the photocenter offset of equation~\eqref{eq:photocentre-pull} where it is blended. Because this pull changes with scan angle, it is not a fixed offset of the origin, so in practice the primary's position is fitted together with the other sources, as in Section~\ref{sec:crowded}. The joint fit estimates the companion's two sky coordinates and both component fluxes. We maximized the likelihood, assuming no informative prior is available.

The $47$ transits of this scan sequence are strongly clustered in scan angle, leaving a gap of approximately $90^\circ$ between sampled directions modulo $180^\circ$, which makes it a hard case for the reconstruction. Such uneven angular coverage occurs over broad regions at low ecliptic latitudes \citep{Holl2023}, making this limitation relevant beyond the particular pointing used here. Its effect on positional precision enters through the scan-angle averages in the information matrix in Appendix~\ref{app: fisher}, equation~\eqref{eq:fisher-matrix}. The inverse of this matrix gives the positional covariance, including the unequal uncertainties and correlations between sky coordinates caused by the angular coverage.

Figure~\ref{fig:scan_to_skyline} illustrates the reconstruction. Of the $47$ windows, $29$ yield a companion detection under the calibrated per-window model comparison. The remaining $18$ are fitted with one source but are still included in the joint window fit. All $29$ companion detections are consistent with the map candidate. The position fitted to these associated detections initializes the joint window fit and is marked by the circle in the inset of panel~(c). The gray band in panel~(b) marks projected separations for which the noise-free blended profile has only one maximum \citep{Penoyre2026}. Such a profile can still support a two-source fit, as for the close blend in Figure~\ref{fig:per_window_flow}, since the comparison uses the full sampled profile rather than counting peaks. Blending nevertheless degrades the positional precision, as quantified in Appendix~\ref{app: fisher}.

The joint fit recovers the companion's position to within approximately $2$ mas of the injected position and gives a flux ratio $\hat{\mathcal{S}}=0.197\pm0.003$, compared with the injected value of $0.2$. The fitted sky offsets are $(189.8\pm1.3,-69.0\pm0.9)$ mas, compared with the injected $(187.9,-68.4)$ mas. The inset of Figure~\ref{fig:scan_to_skyline}c shows the fitted position and its uncertainty contours, together with the starting position.

\begin{figure*}
\centering
\includegraphics[width=\textwidth]{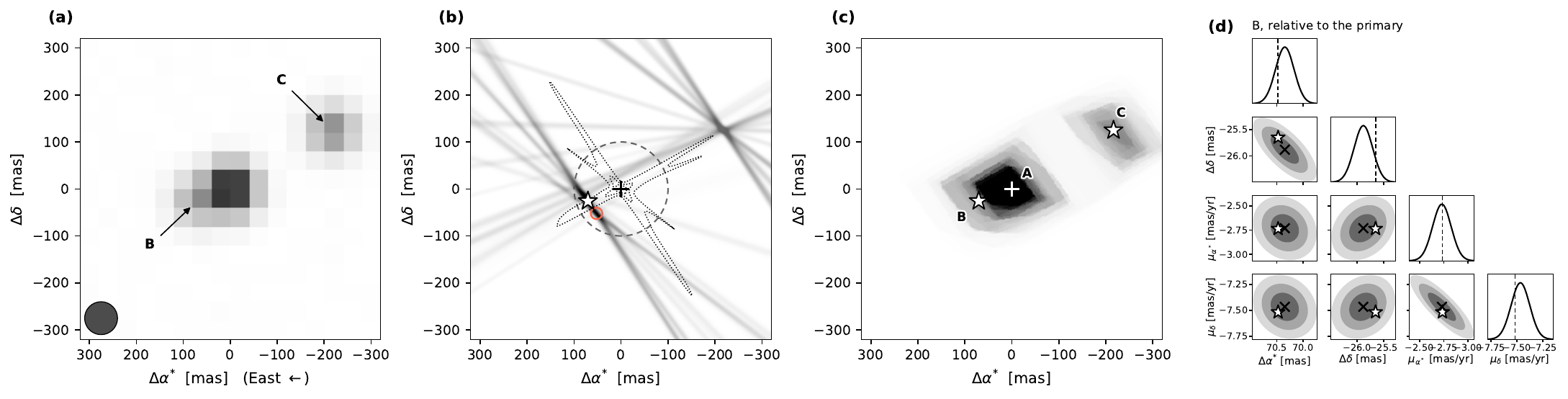}
\caption{Recovery of a close companion in the presence of a contaminating neighbor. The synthetic scene is observed at $47$ GOST-predicted transits under a Gaia-like per-sample noise model for a primary of $G=16$ (Section~\ref{sec:crowded-example}). At the reference epoch, companion B lies at $75$ mas with flux ratio $\mathcal{S}=0.2$ and relative proper motion $8$ mas\,yr$^{-1}$. Neighbor C lies at $250$ mas with flux ratio $\mathcal{S}=0.3$ and is static relative to the primary. \emph{(a)} An illustrative image using a $70$ mas FWHM imaging PSF on $40$ mas pixels. This image does not enter the analysis. \emph{(b)} Triangulation map of the detections, with those assigned to C shown darker and the primary's contributions shown as dotted contours. The open circle marks the map's proposed position for B, and the star marks its injected position. \emph{(c)} A faststack intensity reconstruction from the same $47$ windows \citep[][Appendix~\ref{app:radon}]{Harrison2011}. The plus marks the primary and stars mark the injected positions of B and C. C appears separately, while B contributes a bump near the primary's core. \emph{(d)} Joint uncertainty regions and one-dimensional distributions for B's position and proper motion relative to the fitted primary. Crosses mark fitted values. Stars and dashed lines mark injected values.}
\label{fig:crowded}
\end{figure*}

\subsection{Example: a moving pair}
\label{sec:moving-example}
We next consider a bound companion separated by $200$ mas from the primary at the reference epoch, with a flux ratio of $0.3$. Its orbital motion over the campaign is represented by a relative proper motion of $(3,9)$ mas\,yr$^{-1}$, the linear approximation of Section~\ref{sec:dynamics}. At a distance of $120$ pc, this corresponds to a projected separation of $24$ au and a relative velocity of about $5$ km\,s$^{-1}$, as expected for a pair with a total mass of about one solar mass and an orbital period of about a century. We use the same $47$ transits and the same uniform noise as in the static example. The relative motion produces a displacement of approximately $48$ mas over the sampled campaign. The primary's reference track is again treated as known, while the companion's position and relative proper motion are fitted without informative priors. Figure~\ref{fig:dynamics}a shows the detections and triangulation maps before and after motion compensation.

The fit to the associated along-scan detections gives a relative proper motion of $(3.2\pm1.3,9.0\pm1.1)$ mas\,yr$^{-1}$ and a reference-epoch position within approximately $2$ mas of the injected position. Using this solution to initialize the joint window fit gives $(3.0\pm0.6,8.9\pm0.4)$ mas\,yr$^{-1}$ and a position within approximately $2$ mas of the injected value. The motion compensation in Figure~\ref{fig:dynamics}a uses this joint-fit solution, bringing the detections into closer agreement with a single sinusoid and concentrating the map near the reference-epoch position.

We repeat the analysis over $300$ independent noise realizations. The detection-set fit shows biases of about $1$ mas in each sky coordinate and $0.4$ to $0.5$ mas\,yr$^{-1}$ in each proper-motion component, systematically underestimating the drift. The joint window fit shows no significant bias at the precision of this test, with mean offsets of a few tenths of a milliarcsecond in position. Its scatter is smaller by factors of two to five, depending on the fitted parameter, and agrees with the formal uncertainties. The detection-set uncertainties underestimate the scatter by up to $50\%$ and do not account for the systematic offsets introduced by blending and detection selection, even though the components are separated by twice the LSF width at the reference epoch.

Figure~\ref{fig:dynamics}b shows a chance-aligned source approximately $2$ kpc behind the primary, placed at the same reference-epoch sky offset and assigned an independent proper motion. We now fit five companion parameters: two sky coordinates, two relative proper-motion components, and a parallax difference. The joint window fit gives $\Delta\varpi=-7.4\pm1.0$ mas, compared with the injected value of $-7.8$ mas. Applying the same model to the bound pair in panel~(a) gives $\Delta\varpi=-0.1\pm1.0$ mas, consistent with zero.

\section{Scenes with several sources}
\label{sec:crowded}

We now extend the reconstruction to scenes containing $K$ point sources, including the primary. The joint window model has one LSF column per source, with the likelihood and flux-fitting procedure unchanged. The additional problem is to determine how many sources the observations support and which detections belong to each one. Several sources can produce overlapping tracks in the sinogram, and unrelated detections can form chance crossings in the triangulation map.

\subsection{Multi-object triangulation maps}
\label{sec:multimap}
We build the scene one source at a time, starting from the primary, in three stages.
\begin{enumerate}
    \item Prepare the detection set.
    \begin{itemize}
        \item Fit each window separately (Section~\ref{sec:meas-nnls}).
        \item Identify the primary among the fitted components and remove it from the detection set.
        \item Pool the remaining detections in the triangulation map.
    \end{itemize}
    \item Add sources one at a time.
    \begin{itemize}
        \item Start with a primary-only scene, $K=1$.
        \item Take the highest peak of the map away from the primary as the proposed source. The height of a peak is the vote of the detections whose lines cross there, each weighted by its localization as in equation~\eqref{eq:hough-map}. A candidate must gather a minimum number of crossings, three in this paper, and the search stops when no peak reaches it.
        \item Fit the $K$-source and $(K+1)$-source models to all original windows, the new source started at the proposed position and at rest, all existing sources, including the primary, free to adjust. For a source that moves far over the campaign, as a chance alignment with a large parallax, the sinogram fit of Section~\ref{sec:sinogram} gives a better start.
        \item If the improvement does not exceed the threshold defined below, stop. Otherwise, retain the additional source and set $K\rightarrow K+1$. Remove from the map every detection whose line passes within an error tolerance of any accepted source. The removal uses the source's fitted position at the measurement epoch, as in equation~\eqref{eq: eta_dyn}.
    \end{itemize}
    \item Measure the accepted scene.
    \begin{itemize}
        \item Jointly fit all windows with the accepted sources.
        \item Incorporate external constraints through priors on the relevant source parameters.
    \end{itemize}
\end{enumerate}

Removing detections from the map does not remove their windows from the joint fit, so every comparison uses all windows. An additional source is accepted when
\begin{equation}
\Delta\chi^2\equiv\chi^2_K-\chi^2_{K+1}>\Lambda_{\rm c},
\end{equation}
where $\chi^2_K$ and $\chi^2_{K+1}$ are the minimized values for the two scene models. Searching over candidate positions and motions gives noise more opportunities to resemble an additional source, so the threshold must account for the search rather than only the number of fitted parameters. It also depends on the correlations between nearby trial configurations, which produce similar profiles in the windows.

We calibrate $\Lambda_{\rm c}$ on simulated campaigns that contain only the $K$ sources of the lower-order model. To each campaign we fit the $(K+1)$-source model, with the additional source tried at every position and motion in the field. Because no additional source exists, any improvement in $\chi^2$ comes from noise that the additional source absorbs. The threshold is the value that the largest of these improvements exceeds in $1\%$ of the campaigns. It depends on the scan angles and epochs of the campaign, the instrumental and noise models, and the search bounds.  Under the Gaussian-noise assumptions the threshold can be approximated analytically \citep{Vitells2011}. When the LSF model is correct, the simulated rate at the analytic threshold is $\lesssim1\%$ (Sections~\ref{sec:crowded-example} and~\ref{sec:detection}). When the LSF is not known exactly (Section~\ref{sec:lsf-errors}), the error enters every stage of the search, and the simulation calibrates the whole procedure.

The per-window threshold $\Lambda_{\rm w}$ of Appendix~\ref{app:reduction} follows the same logic for a different fit, testing one additional component within a single window. Both thresholds control only false detections caused by noise. Light that the point-source model does not contain, such as fainter field stars, unresolved galaxies, or structure in the wings of bright neighboring stars, can be fitted as spurious point sources. Gaia's on-board detection rejects spurious events by the shape of their profile and accepts sources only above a flux threshold that sets its faint limit \citep{deBruijne2015}. We likewise limit the companions we claim to a faint magnitude and discuss this and other limitations of the model in Section~\ref{sec:precision}.

\subsection{Example: a close pair with a contaminating neighbor}
\label{sec:crowded-example}
We demonstrate the recovery of a close companion in the presence of a contaminating neighbor. At the reference epoch, companion B lies $75$ mas from the primary at a position angle of $110^\circ$, with a flux ratio of $0.2$. Its relative proper motion is $8$ mas\,yr$^{-1}$, initially perpendicular to the separation vector, producing a displacement of approximately $40$ mas over the campaign. Source C lies $250$ mas from the primary, has a flux ratio of $0.3$, and is static in the primary's frame. It remains within the windows throughout the campaign, and its along-scan position crosses that of B as the scan angle changes.

We use the same strongly clustered sequence of $47$ GOST-predicted transits as in Figures~\ref{fig:scan_to_skyline} and~\ref{fig:dynamics}. The windows carry a Gaia-like per-sample noise model. A binned sample is one along-scan pixel summed over the $12$ across-scan pixels of one CCD crossing. A transit crosses nine CCDs of the astrometric field \citep{Lindegren2018}, which we sum into one window. The noise covariance matrix of window $j$ in equation~\eqref{eq:joint-likelihood} is
\begin{equation}
\mathbf{\Sigma}_j = \operatorname{diag}(\bar{\mathbf{s}}_j) + 9\,(b + r^2)\,\mathbf{I},
\label{eq:noise}
\end{equation}
where $\bar{\mathbf{s}}_j$ is the noise-free signal of the window samples in electrons, $\mathbf{I}$ is the identity matrix, $b=10$ electrons is a quiet background per binned sample per CCD, and $r=4.3$ electrons is the read noise per CCD \citep{Hambly2018}. A primary of $G=16$ yields $3.0\times10^{5}$ electrons per transit, from the $G$ zero point of \citet{Riello2021} and a crossing of $4.42$ s. The background is taken as subtracted exactly, so that only its shot noise remains, and no constant term is fitted (Section~\ref{sec:meas-single}). 

This model sets the sample variances. It is not a calibration of Gaia. The simulation uses the idealized LSF of Section~\ref{sec:static-example}, known exactly, and independent Gaussian noise with that variance, which the fits use as fixed weights. We apply the search and the joint fit of Section~\ref{sec:multimap} to these windows. Figure~\ref{fig:crowded} shows the scene, the triangulation map, an image reconstruction from the same windows, and the fitted motion of B. Figure~\ref{fig:crowded}a illustrates the scene with a $70$ mas FWHM imaging PSF and $40$ mas pixels. Figure~\ref{fig:crowded}c shows a faststack reconstruction of the window samples \citep{Harrison2011}. Its angular source shapes reflect the limited scan-angle coverage and finite sampling of the windows. C appears separately, whereas B is only a bump near the primary's core and not a separate peak, so a search for peaks in this image would not find it.

The search (stages 1 and 2 of Section~\ref{sec:multimap}) uses no prior on the positions of B and C. The per-window fits allow up to three point sources and take the component nearest the window center as the primary. They select one, two, and three sources in $3$, $19$, and $25$ windows, respectively. Removing the primary detections leaves $69$ detections for the triangulation map (Figure~\ref{fig:crowded}b), which proposes B near its injected position, where the image shows no separate peak. The search accepts C first and then B, each with an improvement far above the threshold, $\Lambda_{\rm c}=19.1$,\footnote{Calibrated for the first addition, and used for later additions. In simulated campaigns without a companion, under the noise model of this example, fewer than $1\%$ of the searches exceed it.} after which too few detections remain to propose another candidate. In the final scene fit (stage 3), the primary's position and residual motion are fitted freely, as in the search. For C, we include a Gaussian prior centered on its assumed catalog position, with a width of $5$ mas, to show how an external positional measurement enters the fit. Its width is an illustrative choice for this example.

This fit places B at a separation of $75.0\pm0.1$ mas and a position angle of $110.2^\circ\pm0.2^\circ$, compared with the injected values of $75.0$ mas and $110.0^\circ$. Its relative proper motion is $(-2.7\pm0.1,-7.47\pm0.09)$ mas\,yr$^{-1}$, for an injected $(-2.74,-7.52)$ mas\,yr$^{-1}$ (Figure~\ref{fig:crowded}d). C is recovered at $250.07\pm0.05$ mas and $300.00^\circ\pm0.02^\circ$, and the fitted flux ratios agree with the injected values within their uncertainties of $8\times10^{-4}$ and $3\times10^{-4}$. The primary's fitted offset and residual motion are consistent with zero. Over $600$ noise realizations with random sampling phases, the injected values fall within the quoted $1$, $2$, and $3\sigma$ regions at the expected rates, and the quoted uncertainties of B's four parameters and of both flux ratios match the scatter of the fitted values to within $4\%$.

Although C is well separated from the close pair, it must be included in the model to measure B accurately. Omitting C shifts B's fitted separation by $8$ mas and its position angle by $22^\circ$, with formal uncertainties of $0.1$ mas and $0.1^\circ$. The fit compensates for the missing neighbor by moving B, producing a biased solution with uncertainties far smaller than the displacement.

\section{Precision, model errors and detection limits}
\label{sec:precision}
\subsection{Precision with a known profile}

We first consider the precision attainable when the instrumental profile is known and the sources have been correctly identified. For an isolated source, photon noise gives a positional uncertainty proportional to $f^{-1/2}$ \citep{Mendez2013}, while background noise, detector noise, and calibration errors also contribute in practice.

Overlapping profiles make the secondary's center harder to locate. At small projected separations, $|\rho_\parallel|\ll{\rm FWHM}$, the two profiles merge into a single one. The separation shows only in the excess width and the asymmetry of this profile, that is, in its second and third moments, which grow as $|\rho_\parallel|^{2}$ and $|\rho_\parallel|^{3}$. A shift of the secondary away from the primary broadens the merged profile, but a slightly fainter secondary at the new position restores the original width, so the excess width alone does not reveal the shift. What remains is the asymmetry, which grows as $\mathcal{S}\,|\rho_\parallel|^{3}$, so a small shift $\delta\eta$ changes it by an amount proportional to $\mathcal{S}\,|\rho_\parallel|^{2}\,\delta\eta$. The asymmetry is measured with an error that does not depend on the separation, so the smallest detectable shift $\delta\eta$, and with it the secondary's positional uncertainty, grows as $|\rho_\parallel|^{-2}$. The same asymptotic limit applies to spectroscopic line blending, where it follows from the flux-ratio scaling derived by \citet{Shahaf2026}. In imaging, the signal-to-noise ratio needed to resolve two point sources grows with the same inverse square of their separation \citep{ShahramMilanfar2004}.

\begin{figure*}[t]
\centering
\includegraphics[width=\textwidth]{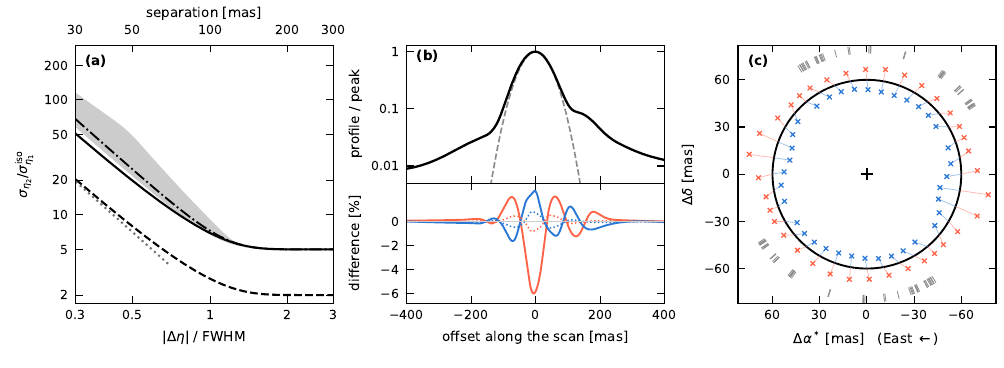}
\caption{Single-window precision of a close companion and the effect of a color error in the LSF. (a) Positional uncertainty of the secondary, normalized by that of the isolated primary, against the projected separation, for Gaussian profiles with $\mathcal{S}=0.2$ (solid) and $0.5$ (dashed), and for the $\mathcal{S}=0.2$ Gaussian sampled as in a Gaia window, at $18$ points spaced by $59$ mas (dash-dotted). The dotted curve shows the small-separation limit for $\mathcal{S}=0.5$. The gray band shows the LSF of panel~(b) at $\mathcal{S}=0.2$, with the color known (lower edge) and fitted in each window (upper edge). (b) Upper panel: the along-scan LSF of a K star, $\nu_{\rm eff}=1.42~\mu{\rm m}^{-1}$, digitized from \citet{Rowell2021} (black), and a Gaussian of FWHM $100$ mas (dashed gray), both normalized to their peak, on a logarithmic scale. Lower panel: the LSF at another color minus that of the K star, in percent of the K star's peak, at the ends of the tabulated range, $\nu_{\rm eff}=1.24$ (red) and $1.72~\mu{\rm m}^{-1}$ (blue), and for a model color in error by $\Delta\nu=-0.03$ (dotted red) and $+0.03~\mu{\rm m}^{-1}$ (dotted blue). (c) Noise-free fits of one campaign for a companion with $\mathcal{S}=0.03$ at $60$ mas from a $G=16$ primary (plus sign), placed at $36$ position angles (circle). Crosses show the fitted positions for a model color in error by $\Delta\nu=-0.01$ (red, too broad) and $+0.01~\mu{\rm m}^{-1}$ (blue, too narrow), and gray ticks show the scan directions.}
\label{fig:blend-precision}
\end{figure*}

Figure~\ref{fig:blend-precision}a shows this loss of precision for two Gaussian profiles with both positions and both fluxes fitted. The uncertainty is normalized by that of the primary fitted in isolation, so it includes both the secondary's lower flux and the effect of blending. At large separations, the curves approach the isolated-secondary limit, and at small separations they follow the $|\rho_\parallel|^{-2}$ law (dotted). To isolate the effect of blending, the calculation uses a finely sampled profile with the same Gaussian noise in every sample. Sampling at the pixel scale of a Gaia window increases the uncertainty further (dash-dotted). Appendix~\ref{app: fisher} gives the calculation for any sampling and noise model and shows how the center measurements combine across scan angles and observing epochs to constrain positions and proper motions.

Positional information therefore persists below the profile FWHM, although recovering it requires increasingly precise measurements. The calculation assumes that the width of the profile is known and that both sources are already identified. 

\subsection{Errors in the line spread function}
\label{sec:lsf-errors}
An incorrect instrumental profile can bias the fitted astrometry and change the inferred number of sources. A profile of the wrong width shifts the fitted positions of blended components. If the adopted profile is too narrow, displaced positive components can reproduce the excess width, creating apparent companions. An overly broad profile can instead absorb part of a close companion's signal and reduce its detectability. 

Gaia's along-scan LSF depends on the color of the source through its effective wavenumber $\nu_{\rm eff}$ \citep{Rowell2021}. We digitized the LSF of \citet[their figure~11]{Rowell2021} at five colors between $\nu_{\rm eff}$ of $1.24$ and $1.72~\mu{\rm m}^{-1}$. To approximate the LSF of a given color, we linearly interpolate in $\nu_{\rm eff}$.\footnote{Linear interpolation between the digitized profiles is accurate to about $1\%$ of the peak. Near $\nu_{\rm eff}=1.42~\mu{\rm m}^{-1}$, the color of a K star, neighboring tabulated profiles differ by about three times this amount, so the digitization resolves the color dependence and suffices for the effects shown here. A more detailed treatment requires the calibrated LSF and will follow with Gaia DR4.} This interpolation serves as a fiducial model of the color dependence. A larger $\nu_{\rm eff}$ corresponds to a bluer source and a narrower profile, with the FWHM decreasing from ${\sim}\,110$ to ${\sim}\,100$ mas. The profile also has broader wings than a Gaussian, with only $80\%$ of its flux within one FWHM of the center, compared with $98\%$ for a Gaussian (Figure~\ref{fig:blend-precision}b, upper panel).

The color of a star is known only to a finite accuracy, and an error in it gives a model LSF of the wrong width and shape. We write the error of the model color as $\Delta\nu=\nu_{\rm eff}^{\rm model}-\nu_{\rm eff}$, so that a negative $\Delta\nu$ gives a model that is too red and therefore too broad. A color error of $0.01~\mu{\rm m}^{-1}$ changes the width of the profile near $\nu_{\rm eff}=1.42~\mu{\rm m}^{-1}$ by about $0.4\%$. The change is not a uniform broadening but a central peak and two pairs of shoulders of alternating sign (Figure~\ref{fig:blend-precision}b, lower panel). The profile of the reddest tabulated color differs from that of the K star by up to $6\%$ of its peak.

Fitting the color together with the positions and fluxes reduces the bias from a color error, at a cost in precision. The gray band in Figure~\ref{fig:blend-precision}a shows this cost for a single window at $\mathcal{S}=0.2$. Its lower edge assumes that the color is known, and its upper edge that each window is fitted with its own color, without information from the photometry. Below $60$ mas, fitting the color approximately doubles the positional uncertainty of the secondary, on average over the sampling phase. A full campaign lies between the two edges, because all its windows share one color, which the photometry also constrains. In the simulations of Section~\ref{sec:detection}, we therefore fix the color and treat its error as a source of bias. We assume that this error is dominated by the calibration of the color, so that it does not depend on magnitude, and take $|\Delta\nu|=0.007~\mu{\rm m}^{-1}$, the accuracy with which the analytic relation of \citet[their equation~3]{Lindegren2021} represents the mean relation between $\nu_{\rm eff}$ and $G_{\rm BP}-G_{\rm RP}$. For a K star, this error corresponds to about $40$ mmag in $G_{\rm BP}-G_{\rm RP}$ and changes the profile by at most $0.2\%$ of its peak. We use this value for the bias in Section~\ref{sec:detection}. The larger errors of $0.01$ and $0.03~\mu{\rm m}^{-1}$ illustrate the effect below and set the recalibrated thresholds.

Figure~\ref{fig:blend-precision}c shows this bias in noise-free fits of one campaign at the $47$ transits of Section~\ref{sec:static-example}, for a companion with $\mathcal{S}=0.03$ at $60$ mas and a model color in error by $0.01~\mu{\rm m}^{-1}$. The fitted companion moves outward when the model is too broad and inward when it is too narrow, by a median of $7$ mas. Next to a primary of $G=16$, this is about three times the quoted uncertainty of the companion's relative position (Section~\ref{sec:detection}), $2.3$ mas. The shift is larger and turns sideways at position angles where the scan directions leave a gap, because few windows there see the pair separated along the scan.

A model that is too narrow can also create companions. The threshold of Section~\ref{sec:multimap} gives $1\%$ false detections when the model is correct. In campaigns without a companion, a model too narrow by $0.01~\mu{\rm m}^{-1}$ raises this rate to ${\sim}\,50\%$ next to a primary of $G=14$ and to ${\sim}\,2\%$ next to $G=16$, but produces none next to $G=18$. A model that is too broad produces no false companions. The threshold must therefore be calibrated under the color error that is expected. We recalibrate $\Lambda_{\rm c}$ with $1000$ campaigns without a companion for each magnitude, fitted with the correct model color and with colors in error by $|\Delta\nu|=0.01$ and $0.03~\mu{\rm m}^{-1}$, and give the cost of these thresholds in Section~\ref{sec:detection}.

\subsection{Detection limits}
\label{sec:detection}

\begin{figure*}[t]
\centering
\includegraphics[width=\textwidth]{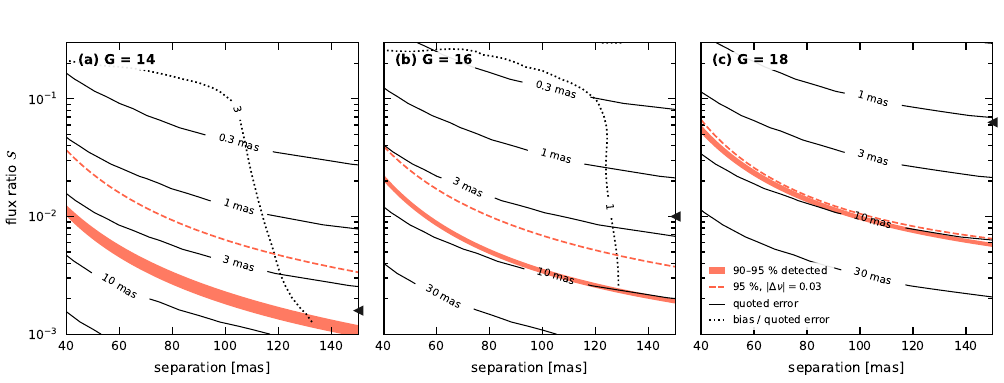}
\caption{Detection limit, positional uncertainty, and color bias of a companion in the plane of separation and flux ratio $\mathcal{S}$, next to a primary of $G=14$ (a), $16$ (b), and $18$ (c). The primary is a K star with $\nu_{\rm eff}=1.42~\mu{\rm m}^{-1}$, observed in $47$ transits with the noise of equation~\eqref{eq:noise} at a background of $b=100$ electrons, and modeled with the LSF digitized from \citet{Rowell2021}. The red band spans the flux ratios at which $90$ to $95\%$ of campaigns detect the companion at a false-detection rate of at most $1\%$, with the color of the LSF known. The dashed curve is the $95\%$ limit with the threshold recalibrated for a color error of $|\Delta\nu|=0.03~\mu{\rm m}^{-1}$. Black contours show the quoted positional uncertainty of the companion. The dotted lines mark where the noise-free bias of the companion's position, for a model color in error by $|\Delta\nu|=0.007~\mu{\rm m}^{-1}$ (Section~\ref{sec:lsf-errors}), equals one and three times that uncertainty, and the bias is larger on the side of closer and fainter companions. The lines are not drawn below the band, and in panel~(c) the bias stays below the uncertainty. The triangle on the right axis marks the flux ratio of a companion of $G=21$.}
\label{fig:detection-plane}
\end{figure*}

We measure the detection limit by simulating campaigns of a primary and a fainter companion. The primary is a K star, with $\nu_{\rm eff}=1.42~\mu{\rm m}^{-1}$ and $G=14$, $16$, or $18$, observed at the $47$ transits of Section~\ref{sec:static-example}. The noise follows equation~\eqref{eq:noise} with $b=100$ electrons, ten times the background of Section~\ref{sec:crowded-example}. Both sources have the color of the primary and are modeled with the LSF of Section~\ref{sec:lsf-errors} at that color, unlike the Gaussian LSF of the examples of Sections~\ref{sec:bp-measure} and~\ref{sec:crowded}. In each campaign, the companion has a random position angle and moves at $2$ mas\,yr$^{-1}$ relative to the primary, perpendicular to the separation.

Each campaign is reduced blindly, from the per-window fits to the joint fit with a free primary. The companion counts as detected when the first candidate of the search exceeds $\Lambda_{\rm c}$. This criterion tests whether the data require a second source, not where that source lies. For this LSF, the analytic approximation of Section~\ref{sec:multimap} gives $\Lambda_{\rm c}=18.8$. None of the null campaigns of Section~\ref{sec:lsf-errors} fitted with the correct color exceeds this value. At nine separations between $40$ and $150$ mas, we simulate $200$ campaigns at each flux ratio, with the same noise realizations at every flux ratio. The limit $\mathcal{S}_{95}$ is the flux ratio at which $95\%$ of the campaigns detect the companion.

Figure~\ref{fig:detection-plane} shows this limit in the plane of separation and flux ratio. Next to a primary of $G=14$, we detect $95\%$ of companions that are $4.7$ mag fainter at a separation of $40$ mas and $7.3$ mag fainter at $150$ mas. The corresponding contrasts are $4.1$ and $6.8$ mag next to $G=16$, and $3.0$ and $5.5$ mag next to $G=18$. A companion of $G=21$ is therefore detected at every simulated separation next to $G=18$, at separations of $60$ mas and more next to $G=16$, and only at $150$ mas next to $G=14$. This magnitude lies near the faint end of Gaia's catalog, at $G\approx20.7$ \citep{GaiaCollaboration2016}. Field stars, background structure, and other light fainter than this are not known to the model (Section~\ref{sec:multimap}), so detections of companions fainter than $G=21$ (Figure~\ref{fig:detection-plane}, triangles) are reliable only where such light can be excluded. Details of the search implementation change these contrasts by about $0.1$ mag.

Thresholds recalibrated for a larger error in the model color (Section~\ref{sec:lsf-errors}) raise the limit (Figure~\ref{fig:detection-plane}, dashed). An error of $0.01~\mu{\rm m}^{-1}$ raises $\mathcal{S}_{95}$ by a factor of $1.1$ to $1.35$ next to $G=14$ and leaves it unchanged next to $G=16$ and $18$. An error of $0.03~\mu{\rm m}^{-1}$ raises it by a factor of $2.5$ to $3.0$ next to $G=14$, $1.7$ to $1.9$ next to $G=16$, and at most $1.13$ next to $G=18$.

The contours of Figure~\ref{fig:detection-plane} give the quoted uncertainty of the companion's position relative to the primary, the square root of the trace of the predicted covariance, taken as the median over eight position angles. We test this uncertainty at the detection limit. At each separation, we take the $200$ campaigns at the flux ratio nearest $\mathcal{S}_{95}$ and compare the offsets of the recovered companions from their injected positions with the covariance predicted for each campaign. The robust width of the offsets is $0.85$ to $1.1$ times the quoted uncertainty, so the uncertainty describes the bulk of the recovered positions. The tail is heavier than Gaussian: $1$ to $12\%$ of the positions lie outside the $3\sigma$ ellipse of the predicted covariance, compared with $0.3\%$ for a Gaussian distribution.

The dotted lines of Figure~\ref{fig:detection-plane} compare the bias for the adopted color error of $|\Delta\nu|=0.007~\mu{\rm m}^{-1}$ (Section~\ref{sec:lsf-errors}) with the quoted uncertainty. We compute the bias without noise, by fitting the windows of one campaign with the model color in error by $\pm0.007~\mu{\rm m}^{-1}$, with the primary free and the companion started at its true position. The bias is the median offset of the fitted companion over eight position angles, for the sign of $\Delta\nu$ that gives the larger offset. Next to a primary of $G=14$, the bias exceeds the quoted uncertainty over the whole detected region, typically by a factor of three, and amounts to $21$ to $26$ mas at the detection limit, at separations of $40$ to $80$ mas. Next to $G=16$, it exceeds the quoted uncertainty over about two thirds of the detected region, by up to a factor of $2.6$ and by $11$ to $14$ mas at the detection limit, and falls below it beyond about $125$ mas. Next to $G=18$, it stays below the quoted uncertainty, at most ${\sim}\,0.8$ times it. The same color error therefore costs more next to a brighter primary, for two reasons. The companions detected there are fainter relative to the primary, so the misfit of the primary's profile shifts them by more, and the larger number of photons makes the quoted uncertainty smaller, so the same shift is a larger fraction of it.

The companion's own light also enters the measured color, because a neighbor within about $1$ arcsec is blended into the BP and RP fluxes \citep[section~9.3]{Riello2021}. We estimate that a companion redder than the primary by $1$ mag in $G_{\rm BP}-G_{\rm RP}$ shifts the color of the pair by about $1.1\,\mathcal{S}$ mag, which is $11$ mmag at $\mathcal{S}=0.01$ and corresponds to $\Delta\nu\approx0.0018~\mu{\rm m}^{-1}$. This shift is about a quarter of the adopted color error, and it makes the model profile of the primary too red and therefore too broad. The simulations do not include this effect. Fitting the color with the scene reduces it at a cost in precision (Section~\ref{sec:lsf-errors}), whereas the results of Figure~\ref{fig:detection-plane} assume that the color of the primary is known.

\section{Discussion}
\label{sec:discussion}

Gaia's repeated scans contain information about the local stellar scene that a single-source position does not capture. The triangulation map turns the per-window detections into candidate configurations, and the joint fit to the original windows measures them, also in windows in which the components were not detected separately. Relative motion and contamination enter one scene model rather than the error budget. The map is not an image. An intensity image describes the scene without a model, including extended emission and structure that a point-source model does not anticipate, and a stack reveals sources too faint for any single window. The scene model instead names a small number of point sources and measures their positions, fluxes and motions through a forward model.

The main simplification is that the scene holds a few point sources, in each window and on the sky. This is the sparsity assumption of the per-window problem, for which convex and greedy routes exist. Total-variation and $\ell_1$ minimization recover well-separated point sources exactly from band-limited data \citep{Candes2012,Bendory2014}, and for non-negative sources the separation condition can be dropped \citep{Bendory2015}. For non-negative point sources on a dictionary of unit-sum profile columns, the $\ell_1$ norm of the flux vector is the total model flux, which the fit already holds close to the counts in the window, so the non-negativity of non-negative least squares supplies the sparsity with no parameter to tune. An $\ell_1$ penalty becomes useful for signed atoms, such as the wavelets or shearlets discussed below. 

Greedy decompositions \citep{Mallat1993}, of which CLEAN is the astronomical form \citep{Hogbom1974}, select components one at a time and inherit the order dependence of the first choice. We instead fit a small, explicit number of sources with calibrated thresholds, which places the problem in the hypothesis-testing frame of two-point resolution \citep{ShahramMilanfar2004}. When the scene is harder, with more sources than the windows resolve or with light between them, the fit needs further constraints. The first limit is the per-window reduction, which is calibrated here for up to three sources per window (Appendix~\ref{app:reduction}). A window that holds more sources returns the closest pair as one component, with an uncertainty that does not include the blend, and the map and the first joint fits inherit that bias. The joint fit of all windows has no such cap, so crowded fields need a per-window reduction calibrated for more sources, or priors on the scene and external data. The same construction reaches beyond point sources. The LSF columns are atoms, and a scene is their linear combination. Atoms that describe extended light, such as wavelets, shearlets or curvelets projected onto the scan direction, would let the same fit describe resolved sources, provided that the extended light is also sparse in these atoms, at the price of a larger dictionary and its own calibration.

The method assumes a known instrumental profile and independent Gaussian noise of known variance. An error in the profile, in its color or its time dependence, can create components or bias them. Section~\ref{sec:precision} quantifies the cost for this error. The per-window threshold $\Lambda_{\rm w}$ is calibrated for a known kernel, so a color error lets false detections through it. They are stopped at the campaign level by triangulation map crossing limits and the $\Lambda_{\rm c}$ threshold, at a cost in sensitivity. The limits of Section~\ref{sec:detection} are measured on one scan law, that of the Pleiades with its $90^\circ$ gap. Another law changes the transit count, the direction coverage and the calibration of $\Lambda_{\rm c}$, so the limits need their own calibration for each law. Correlated noise, a residual background and the window boundaries also enter the significance and the precision, and application to Gaia windows requires the calibration of the source-count comparisons and of the reported uncertainties under real conditions. Sources brighter than $G\simeq13$ are observed in two-dimensional windows, whose point-spread function depends on the across-scan image motion during integration, whereas the LSF does not \citep{Rowell2021}.

The method can extend the census of companions at close projected separations and give relative motions in fields where crowding disturbs the astrometry. Modeling the neighbors would show whether an anomalous Gaia solution reflects an unmodeled source. External information enters the fit as priors, so a companion found by another instrument can be searched for in Gaia's windows, and an orbit constrained independently can guide a search for companion light around a compact-object candidate \citep{Panuzzo2024}. The color of the profile is a parameter of the model, and Section~\ref{sec:lsf-errors} gives its cost when it is fixed. Fitting the color with the scene would let the windows refine the color of each component. Two extensions follow from the model itself. Atoms for extended light would admit resolved sources, and fluxes that vary with time would admit variable stars and binaries with a Cepheid, through a joint model of brightness variations and orbital motion.

\section*{Data availability}
This work uses simulated observations only. The scan-law forecasts are the public GOST predictions for the fields named in the text. The code that implements the method is available at \url{https://github.com/saharsh1/scanlib}.

\begin{acknowledgements}
S. Shahaf and H. W. Rix acknowledge support from the European Research Council for the ERC Advanced Grant [101054731]. S. Shahaf thanks Eran Ofek and Tamar Faran for the helpful discussions, and Stav Bar-Sheshet for discussions on multi-start optimization and non-negative least squares.
\end{acknowledgements}




\bibliographystyle{aa}
\bibliography{references}

\begin{appendix}
\nolinenumbers
\section{The per-window reduction}
\label{app:reduction}
\subsection{Initialization}
Section~\ref{sec:measurement} reduces each window to its detection set $D_{j}$. The reduction fits models with increasing numbers of point sources, represented by a known LSF, and compares the fitted models to determine how many sources the window supports. Each source has a fitted position and a non-negative flux. Under independent additive Gaussian noise with known, fixed variance $\sigma_n^2$ common to all samples, maximum-likelihood estimation is equivalent to $\chi^2$ minimization. At any trial set of positions, the fluxes are obtained by a non-negative linear solve and profiled out, leaving only the positions to be searched. These are allowed to vary over the sampled range extended by one LSF FWHM on either side. The procedure applies to general source counts. The demonstrations in this paper focus primarily on binary scenes.

The non-negative solve of equation~\eqref{eq:nnls} is first run on the over-complete dictionary, with one LSF column per node of a grid spaced at one eighth of an along-scan pixel ($7.4$~mas). No detection threshold is applied at this stage. The active-set solver returns zero flux for inactive nodes. To exclude numerical residue, we retain as active only nodes with fitted flux greater than $10^{-9}$ times that of the brightest node. The active nodes are sorted by position and grouped into clusters, with a new cluster starting whenever the gap between consecutive active nodes exceeds one along-scan pixel ($59$~mas). Clusters are ranked by the sum of their active-node fluxes. These nodes and clusters provide starting positions for the fits below, rather than being interpreted as detected sources. A window with no active node is recorded as containing no source and contributes an empty detection set.

The sparse grid solution does not directly determine the source count and position. A single source can populate several nodes, while nearby sources can occupy the same cluster. We therefore use it only to initialize fits with increasing numbers of sources. Each fit refines the positions and fluxes jointly, and an additional component is retained only when its improvement over the simpler model exceeds a simulation-calibrated threshold. The sequence begins with one source and stops when the next component is not supported.

\subsection{Position refinement and source count}

Each fit needs starting positions, which come from the ranked clusters and nodes of Appendix~\ref{app:reduction}.1. The one-source fit starts at the brightest node of the brightest cluster. The two-source fit starts from three configurations, one for each way in which the second source can appear in the sparse solution. If the two sources are blended, they occupy one cluster, and the first start takes the two brightest nodes of the brightest cluster. If they are resolved, they occupy two clusters, and the second start takes the brightest node of each of the two brightest clusters. If the second source is too faint to receive a node, the third start keeps the fitted one-source position and adds the grid node at which an additional non-negative LSF reduces $\chi^2$ the most. Starting positions closer than one grid step are separated by one step. All positions are then refined jointly from each start, and the fit with the lowest $\chi^2$ is retained.

Fits with $K$ sources start in the same way. The cluster starts take $K$ nodes from the $K$ brightest clusters, in every combination that allows a cluster to be chosen more than once, with a cluster chosen $m$ times contributing its $m$ brightest nodes. The residual start keeps the fitted positions of the $(K-1)$-source fit and adds the node at which an additional source reduces $\chi^2$ the most. This start also makes the sequence of fits nested. The previous fluxes with a zero flux for the added source are a feasible point of the $K$-source fit, so its $\chi^2$ is never larger than that of the $(K-1)$-source fit, and the improvement $\Delta\chi^2$ is never negative.

An additional source is accepted only if $\Delta\chi^2$ exceeds the calibrated threshold $\Lambda_{\rm w}$. For the comparison between $K-1$ and $K$ sources, the full fitting procedure is applied to simulated windows containing $K-1$ sources, with noise drawn from the assumed noise model. The threshold is the $99$th percentile of the resulting improvements, corresponding to an approximately $1\%$ false-detection rate per window over that calibration ensemble. A standard $\chi^2$ reference distribution is not generally applicable because the added flux is constrained to be non-negative and its position is undefined when its flux is zero. The simulations include the search over positions as actually performed.

For the adopted LSF, $18$-sample windows, noise model, and fitting procedure, the calibrated thresholds are $11.1$ for adding a second source and $10.8$ for adding a third. We therefore adopt $\Lambda_{\rm w}=11$ as a nominal threshold for these two comparisons. These per-window thresholds do not depend on the number or scan angles of the other observations in the campaign. Other source counts or changes to the window configuration, noise model, or search procedure require calibration for the corresponding case. The sequence stops when an additional source is not supported.

\section{Astrometric precision}
\label{app: fisher}
This appendix describes how along-scan precision, scan angles, and observing epochs determine the precision of source positions and proper motions. We first consider the positional information in an individual window, including the effect of blending, and then combine along-scan position measurements across a campaign. These calculations provide simplified estimates for interpreting the results. The measurements reported in the main text instead come from a joint fit to the original window samples.

\subsection{From window samples to source positions}

This subsection shows when the window samples can be replaced by fitted source centers and their covariance for positional inference. We use a local approximation around a fitted solution and allow the source fluxes to vary.

For a fixed number of sources, the expected flux in sample $p$ is $m_p=\sum_k f_k\mathcal{P}(x_p-\eta_k)$, where $f_k$ and $\eta_k$ are the source fluxes and along-scan centers. We assume that the instrumental profile $\mathcal{P}$ is known and denote the fitted fluxes and centers by $\hat f_k$ and $\hat\eta_k$. To describe the positional information near this solution, we consider small changes in both quantities, written as $\delta f_k=f_k-\hat f_k$ and $\delta\eta_k=\eta_k-\hat\eta_k$. Expanding the profile to first order in the center displacement gives
\begin{equation}
\mathcal{P}(x-\eta_k)
\approx
\mathcal{P}(x-\hat\eta_k)
-\delta\eta_k\,\mathcal{P}'(x-\hat\eta_k).
\end{equation}
A flux change multiplies the profile itself, while a center displacement changes it through its derivative. Keeping both contributions gives
\begin{equation}
\delta m_p
\approx
\sum_k\left[
\delta f_k\,\mathcal{P}_k(x_p)
-\hat f_k\,\delta\eta_k\,\mathcal{P}'_k(x_p)
\right],
\label{eq:app-model-change}
\end{equation}
where $\mathcal{P}_k(x)=\mathcal{P}(x-\hat\eta_k)$. We retain only first-order terms in the parameter deviations.

We assume Gaussian noise with covariance $\mathbf{\Sigma}$, allowing for correlations between the window samples. Substituting the first-order model into $-2\ln\mathcal{L}=(\vec{s}-\vec{m})^{\intercal}\mathbf{\Sigma}^{-1}(\vec{s}-\vec{m})$ gives a constant, a term linear in the deviations and a quadratic term. The linear term is proportional to the gradient of $\chi^2$, which vanishes at a best fit with positive fluxes. What remains is a local quadratic log-likelihood in the flux and position deviations,
\begin{equation}
-2\ln\mathcal{L}
\approx \mathrm{const}
+
\begin{pmatrix}
\delta\boldsymbol{\eta}\\
\delta\boldsymbol{f}
\end{pmatrix}^{\intercal}
\begin{pmatrix}
\mathcal{I}_{\eta\eta} & \mathcal{I}_{\eta f}\\
\mathcal{I}_{f\eta} & \mathcal{I}_{ff}
\end{pmatrix}
\begin{pmatrix}
\delta\boldsymbol{\eta}\\
\delta\boldsymbol{f}
\end{pmatrix}.
\end{equation}
The information matrix is $\mathcal{I}=\mathbf{J}^{\intercal}\mathbf{\Sigma}^{-1}\mathbf{J}$, where the columns of $\mathbf{J}$ are the derivatives of the model in equation~\eqref{eq:app-model-change} with respect to the positions and fluxes. It describes how strongly the window constrains changes in positions and fluxes. For independent sample errors with a common flux uncertainty $\sigma_n$, its blocks follow from ordinary inner products of the profiles and their derivatives,
\begin{equation*}
\begin{aligned}
(\mathcal{I}_{\eta\eta})_{k\ell}
&\propto \hat f_k\hat f_\ell\,\mathcal{P}'_k\cdot\mathcal{P}'_\ell,\\
(\mathcal{I}_{\eta f})_{k\ell}
&\propto -\hat f_k\,\mathcal{P}'_k\cdot\mathcal{P}_\ell,\\
(\mathcal{I}_{ff})_{k\ell}
&\propto \mathcal{P}_k\cdot\mathcal{P}_\ell.
\end{aligned}
\end{equation*}
All three expressions have the same factor $\sigma_n^{-2}$. The dot denotes the ordinary inner product over the window samples, and $\mathcal{I}_{f\eta}=\mathcal{I}_{\eta f}^{\intercal}$. The cross block measures the overlap between the effects of changing a center and changing a flux.

This overlap allows a flux adjustment to compensate for part of a positional change. At each choice of center displacements, we fit the fluxes by minimizing the quadratic negative log-likelihood. We differentiate with respect to the flux deviations $\delta\boldsymbol f$, set the derivative to zero, and solve for the flux adjustment to obtain
\begin{equation}
\delta\boldsymbol{f}_{\rm best}
=
-\mathcal{I}_{ff}^{-1}
\mathcal{I}_{f\eta}\,
\delta\boldsymbol{\eta}.
\end{equation}
Substituting this adjustment into the joint likelihood leaves the profiled positional likelihood,
\begin{equation}
-2\ln\mathcal{L}(\boldsymbol{\eta})
\approx \mathrm{const}
+\delta\boldsymbol{\eta}^{\intercal}
\mathbf{C}_{\eta}^{-1}
\delta\boldsymbol{\eta},
\label{eq:profiled-position-likelihood}
\end{equation}
with 
\begin{equation*}
\mathbf{C}_{\eta}^{-1}
=
\mathcal{I}_{\eta\eta}
-\mathcal{I}_{\eta f}
\mathcal{I}_{ff}^{-1}
\mathcal{I}_{f\eta}.
\end{equation*}
The subtraction term removes the positional information that can be reproduced by changing the fluxes. The remaining information determines the precision of the centers when the fluxes are fitted as well.

Here, $\mathbf{C}_{\eta}$ is the covariance matrix of the fitted centers after fitting the fluxes. Near an identifiable solution with a fixed source count and positive fitted fluxes, the centers and their covariance retain the local positional likelihood of the window. This is the uncertainty calculation underlying the per-window fits in Section~\ref{sec:meas-nnls}. It does not yet include the shared-flux constraint across transits. We next use this covariance to describe the uncertainty in the companion's fitted position across a time series.

\subsection{Triangulation error}
The calculation in B.1 also applies to a time series in which the source fluxes are shared across transits. For $K$ sources observed in $n$ transits, we collect the $nK$ along-scan centers into one vector $\boldsymbol\eta$ and fit them jointly with the $K$ shared fluxes. Profiling over these fluxes gives the same quadratic positional likelihood as in B.1. The resulting covariance $\mathbf C_\eta$ now includes correlations between centers in different transits.

An astrometric model with parameters $\boldsymbol\theta$ predicts the centers $\boldsymbol\eta(\boldsymbol\theta)$. Substituting these predictions gives
\begin{equation}
-2\ln\mathcal L(\boldsymbol\theta)
\approx\mathrm{const}
+[\hat{\boldsymbol\eta}-\boldsymbol\eta(\boldsymbol\theta)]^{\intercal}
\mathbf C_\eta^{-1}
[\hat{\boldsymbol\eta}-\boldsymbol\eta(\boldsymbol\theta)].
\label{eq:joint-centre-likelihood}
\end{equation}
Here $\hat{\boldsymbol\eta}$ denotes the centers fitted jointly with shared fluxes, before imposing the astrometric model. The source model and associations across transits are held fixed. Within the quadratic approximation, the jointly fitted centers and their full covariance retain the profiled positional likelihood.

For a simple estimate of triangulation precision, we now allow the fluxes to be fitted independently in each transit and retain only measurements in which the two sources are resolved. This simplifies the calculation but can discard information supplied by the shared fluxes. With independent window noise, fitting the windows separately also makes their center estimates independent.

We now consider the fitted center of the secondary in the adopted astrometric reference frame. Here both source positions are fitted. Choosing the coordinate origin does not fix the primary's position. Each transit supplies a center estimate and a positional uncertainty obtained from the corresponding diagonal entry of the fitted covariance matrix. For the simple calculation below, we assume a common uncertainty $\sigma_\eta$ for the retained transits. All sums and averages below refer to these transits. For a secondary at fixed sky coordinates $(\Delta\alpha^{*},\Delta\delta)$ relative to the adopted origin, its along-scan center $\eta_j$ in transit $j$, observed at scan angle $\psi_j$, is given by equation~\eqref{eq:projection}.
Combining the $n$ independent center measurements $\hat\eta_j$ gives, up to an additive constant,
\begin{equation}
-2\ln\mathcal{L}=\frac{1}{\sigma_\eta^{2}}\sum_{j=1}^{n}
\left(\hat\eta_j-\Delta\alpha^{*}\sin\psi_j-\Delta\delta\cos\psi_j\right)^{2},
\label{eq:app-static-chisq}
\end{equation}
which is quadratic in the two unknowns. The matrix of second derivatives of $-\ln\mathcal{L}$ with respect to $(\Delta\alpha^{*},\Delta\delta)$ gives the campaign's information matrix,
\begin{equation}
\mathcal{I}=\frac{n}{\sigma_\eta^{2}}
\begin{pmatrix}
\langle\sin^{2}\psi\rangle & \langle\sin\psi\cos\psi\rangle\\[2pt]
\langle\sin\psi\cos\psi\rangle & \langle\cos^{2}\psi\rangle
\end{pmatrix} \longrightarrow \frac{1}{2}\,\frac{n}{\sigma_{\eta}^{2}}\;\mathbf{1},
\label{eq:fisher-matrix}
\end{equation}
where $\langle\,\cdot\,\rangle$ denotes the average over the $n$ retained transits in which the sources are resolved. If their scan angles are uniformly distributed, the information matrix becomes proportional to the identity, as indicated above. In practice, the angular coverage depends on both the scanning law and which transits are retained.

The positional errors are obtained by inverting the matrix in equation~\eqref{eq:fisher-matrix}. In the homogeneous limit they are similar in size,
\begin{equation}
\sigma_{\Delta\alpha^{*}}\simeq\sigma_{\Delta\delta}\simeq
\sqrt{\frac{2}{n}}\,\sigma_\eta.
\label{eq:sigma-static}
\end{equation}
The two coordinate errors are uncorrelated in this limit. For illustration, $n=47$ resolved measurements with $\sigma_\eta=5$~mas and approximately uniform scan angles would give an uncertainty of about $1$~mas in each coordinate. This uses the same number of transits as the fiducial campaign in Section~\ref{sec:dynamics}. The joint forward fit also uses blended observations, so this estimate provides only an approximate comparison with its recovered precision.

\subsection{Proper motion error}
\label{app:propermotion}

We now allow the secondary to move with a constant proper motion in the adopted reference frame. We assume a bound pair, so the two sources share one parallax and their relative position carries no parallax term. The positional model in equation~\eqref{eq:projection} then gains the terms $\mu_{\alpha^*}\tau\sin\psi$ and $\mu_\delta\tau\cos\psi$, where $\tau=t-t_0$ is the time from a reference epoch. We retain independent Gaussian measurement errors but allow the along-scan center uncertainty $\sigma_{\eta,j}$ to differ between transits.

For blended measurements, the same calculation provides an approximate uncertainty estimate when fitting both sources yields approximately unbiased secondary centers with Gaussian errors that are independent between transits. The effect of blending enters through $\sigma_{\eta,j}$, obtained from the corresponding diagonal entry of the fitted center covariance in each window. This approximation does not account for systematic shifts in the fitted centers or multiple distinct fit solutions.

We assign each transit a normalized inverse-variance weight $w_j=\sigma_{\eta,j}^{-2}/\sum_{j'}\sigma_{\eta,j'}^{-2}$, so that $\sum_j w_j=1$, and summarize the scan angles and observing epochs through
\begin{equation}
z_m
\equiv
\left\langle\tau^m e^{2\mathrm{i}\psi}\right\rangle
=
\sum_j w_j\,\tau_j^m e^{2\mathrm{i}\psi_j},
\label{eq:zk}
\end{equation}
where $m=0,1,2$. The factor of two accounts for the equivalent angular information supplied by scan directions separated by $\pi$. Throughout this subsection, angle brackets denote inverse-variance-weighted averages over the measurements used. More precise measurements therefore contribute more strongly. For equal uncertainties, $w_j=1/n$, and these averages reduce to those used in the preceding subsection.

Using these weighted averages, we define
\begin{equation*}
\mathrm{Q}_m \equiv
\frac{n}{2\overline v}
\begin{pmatrix}
\langle\tau^m\rangle-\operatorname{Re}z_m &
\operatorname{Im}z_m\\
\operatorname{Im}z_m &
\langle\tau^m\rangle+\operatorname{Re}z_m
\end{pmatrix},
\end{equation*}
where $\overline v=n/\sum_j\sigma_{\eta,j}^{-2}$ is the harmonic mean of the measurement variances, so that $n/\overline v=\sum_j\sigma_{\eta,j}^{-2}$. For equal uncertainties, $\overline v=\sigma_\eta^2$. These matrices include the absolute measurement precision. The matrix $\mathrm{Q}_0$ describes the weighted angular coverage, while $\mathrm{Q}_1$ and $\mathrm{Q}_2$ also include the observing epochs. For the parameter vector $\boldsymbol{\theta}= (\Delta\alpha^*,\Delta\delta,\mu_{\alpha^*},\mu_\delta)$, the information matrix is
\begin{equation}
\mathcal{I}
=
\begin{pmatrix}
\mathrm{Q}_0 & \mathrm{Q}_1\\
\mathrm{Q}_1 & \mathrm{Q}_2
\end{pmatrix}.
\end{equation}
Position and proper motion are coupled through $\mathrm{Q}_1$. The position and proper-motion blocks of the covariance matrix are
\begin{equation}
\begin{aligned}
\operatorname{Cov}(\Delta\alpha^*,\Delta\delta)
&=
\left(
\mathrm{Q}_0-\mathrm{Q}_1\mathrm{Q}_2^{-1}\mathrm{Q}_1
\right)^{-1},\\
\operatorname{Cov}(\mu_{\alpha^*},\mu_\delta)
&=
\left(
\mathrm{Q}_2-\mathrm{Q}_1\mathrm{Q}_0^{-1}\mathrm{Q}_1
\right)^{-1}.
\end{aligned}
\end{equation}
The remaining covariance terms can be block-inverted as well, but are not shown here explicitly.

We choose the reference epoch as the weighted mean, $t_0=\sum_j w_j t_j$, so that $\langle\tau\rangle=0$. For a simple limiting case, we further assume that the weighted distribution of scan angles is approximately uniform and does not depend on epoch. Then $\mathrm{Q}_1\simeq0$ and $\mathrm{Q}_2\simeq\langle\tau^2\rangle\mathrm{Q}_0$, so position and proper motion approximately decouple. The uncertainty in either proper-motion component becomes
\begin{equation}
\sigma_\mu
\simeq
\sqrt{\frac{2\overline v}{n\langle\tau^2\rangle}}
\approx
\frac{5}{T}\sqrt{\frac{\overline v}{n}},
\label{eq:sigma-mu}
\end{equation}
where the last expression assumes approximately uniform temporal coverage after weighting over a baseline $T$, giving $\langle\tau^2\rangle\simeq T^2/12$. For equal measurement uncertainties, this reduces to $\sigma_\mu\simeq5\sigma_\eta/(T\sqrt{n})$. These simplified conditions provide a reference estimate. The full covariance expressions retain the actual angular coverage, observing epochs, and measurement uncertainties.

The calculation above fits the secondary's position and constant proper motion. For a chance alignment, the two sources have different parallaxes, and the differential parallax $\Delta\varpi$ adds the along-scan parallax factor $f_\varpi$ of equation~\eqref{eq:sinogram-dynamics-extended} as a fifth column, next to $\sin\psi$, $\cos\psi$, $\tau\sin\psi$ and $\tau\cos\psi$. The information matrix then gains one row and one column, built in the same way. Fitting a constant acceleration adds the columns $\tfrac{1}{2}\tau^2\sin\psi$ and $\tfrac{1}{2}\tau^2\cos\psi$. These additional parameters must be included when constructing and inverting the information matrix. Their correlations with position and proper motion can increase the uncertainties, so the proper-motion precision must then be obtained from the expanded covariance matrix.

\subsection{Approximate sensitivity to relative motion}
We estimate the significance of a companion's relative orbital motion from the difference between the two fitted proper motions. We assume a circular orbit observed over a sufficiently short arc that the fitted relative motion approximates its instantaneous value.

The projection factor $f_{\rm proj}$ is the fraction of the relative orbital velocity that lies in the plane of the sky. It depends on orbital orientation and phase, with $0\leq f_{\rm proj}\leq1$. For a circular orbit with total mass $M$, period $P$, and system parallax $\varpi$, the instantaneous relative proper-motion amplitude is
\begin{equation}
\mu_{\rm rel}
=
2\pi f_{\rm proj}
\left(\frac{\varpi}{\mathrm{mas}}\right)
\left(\frac{M}{M_\odot}\right)^{1/3}
\left(\frac{P}{\mathrm{yr}}\right)^{-1/3}
\,\mathrm{mas\,yr^{-1}}.
\label{eq:relative-motion-scale}
\end{equation}

For this illustrative estimate, we assume independent errors in the two fitted proper motions and denote the faint companion's per-component uncertainty by $\sigma_\mu$. Using equation~\eqref{eq:sigma-mu}, we estimate the significance of the relative motion as
\begin{equation}
\mathrm{SNR}_{\mu}
\approx\frac{\mu_{\rm rel}}{\kappa\sigma_\mu}
\approx\frac{\mu_{\rm rel}T}{5\kappa}
\sqrt{\frac{n}{\overline v}}.
\label{eq:motion-significance-period}
\end{equation}
Here $\overline v$ refers to the companion's center measurements and $T$ is the observing baseline, for which we adopt $5.5$ yr for Gaia DR4. We adopt $\kappa=1$ as the fiducial assumption, corresponding to an uncertainty dominated by the faint companion. If the two fitted motions have similar uncertainties, $\kappa\approx\sqrt{2}$.

For the most favorable projection, $f_{\rm proj}=1$, and the fiducial $\kappa=1$, substituting equation~\eqref{eq:relative-motion-scale} and $T=5.5$ yr gives
\begin{equation}
\mathrm{SNR}_{\mu}\approx
6.9\,\sqrt{n}
\left(\frac{\overline v}{\mathrm{mas}^{2}}\right)^{-1/2}
\left(\frac{\varpi}{\mathrm{mas}}\right)
\left(\frac{M}{M_\odot}\right)^{1/3}
\left(\frac{P}{\mathrm{yr}}\right)^{-1/3}.
\label{eq:motion-significance-units}
\end{equation}

At fixed $\kappa\sigma_\mu$ and projection factor, with the other quantities held constant, the motion significance decreases with orbital period as $P^{-1/3}$. Shorter periods give faster orbital motion, but at fixed total mass and distance they also imply smaller angular orbits. The resulting smaller projected separations can increase blending and the uncertainties in the fitted motions, offsetting the faster motion. A companion may therefore remain detectable even when its relative motion cannot be measured precisely.

\section{Triangulation maps and the Radon transform}
\label{app:radon}

\subsection{The triangulation map as a positional statistic}
\label{app:map-statistic}
The triangulation map in equation~\eqref{eq:hough-map} measures the agreement between the sky positions allowed by individual detections. Each detection contributes a band around its allowed sky line, with a Gaussian profile perpendicular to that line. This profile has unit integral and a width set by the along-scan uncertainty, so less precise detections contribute broader bands with lower peaks. The fitted source flux does not enter the weight. Adding these contributions highlights candidate positions without first deciding which detections belong to the same source. This construction is a form of Hough accumulation \citep{DudaHart1972}. The resulting map is a positional statistic, not a likelihood surface. For independent detections known to belong to one source, the positional likelihood would instead be obtained by multiplying their Gaussian factors.

\subsection{Intensity profiles and Radon inversion}
For a static scene, an ideal scan measures a one-dimensional profile obtained by integrating the surface brightness perpendicular to the scan coordinate. These profiles, collected over scan angle, form the Radon transform of the scene. With sufficient angular coverage, the two-dimensional brightness distribution can be recovered by filtered backprojection \citep{Kak2001}. Gaia windows depart from these ideal projections through instrumental blurring, finite window size, and incomplete angular coverage. Our triangulation map uses the same projection geometry, but combines fitted source positions rather than reconstructing brightness from the intensity profiles.

The intensity profiles arranged by scan angle form an intensity sinogram. For a static scene of point sources, the expected signal is
\begin{equation}
s(\eta\mid\psi)
=
\sum_k f_k\,\mathcal{P}\!\left[
\eta-\rho_k\cos(\psi-\varphi_k)
\right],
\end{equation}
where $f_k$ is the flux of source $k$, and $(\rho_k,\varphi_k)$ specifies its position relative to the window reference point. For this idealization, the normalized along-scan instrumental profile $\mathcal{P}$ is taken as known and independent of position, scan angle, and source color.

Let $I(\Delta\alpha^*,\Delta\delta)$ denote the sky image including instrumental blurring. Assuming that the scan profiles are projections of this image, the projection-slice theorem relates their Fourier transforms to that of the image \citep{Kak2001},
\begin{equation}
\hat{s}(\nu\mid\psi)
=
\hat{I}(\nu\sin\psi,\nu\cos\psi),
\end{equation}
where $\nu$ is the spatial frequency conjugate to $\eta$. Each scan samples a line through the origin in the image's Fourier plane.

The image can be reconstructed by filtering each scan profile and backprojecting it along the corresponding sky lines,
\begin{equation}
    I(\Delta\alpha^*, \Delta\delta)\approx \sum_j   q\left(\Delta\alpha^{*}\sin\psi_j + \Delta\delta\cos\psi_j\,|\,\psi_j\right) \,\delta\psi_j,
    \label{eq:fbp}
\end{equation}
where $q$ is the ramp-filtered scan, defined in the Fourier domain by $\hat{q}(\nu\mid\psi)\equiv|\nu|\,\hat{s}(\nu\mid\psi)$ and $\delta\psi_j$ is the angular weight assigned to the scan. We resample the filtered profiles onto a finer along-scan grid to interpolate their values at the projected positions of image pixels.

For irregular angular sampling, each weight $\delta\psi_j$ represents the angular interval covered by its scan. Scans separated by $180^\circ$ sample the same projection with the coordinate $\eta$ reversed, although their measurement noise can be independent. We therefore fold the angles into $[0,\pi)$, reversing $\eta$ whenever an angle is shifted by $\pi$, and sort them. For distinct folded angles, each scan receives half the sum of the angular gaps to its preceding and following neighbors, with wraparound at $0$ and $\pi$. For $N$ uniformly spaced directions, this gives $\delta\psi_j=\pi/N$.

\begin{figure*}[p]
\centering
\includegraphics[width=0.92\textwidth]{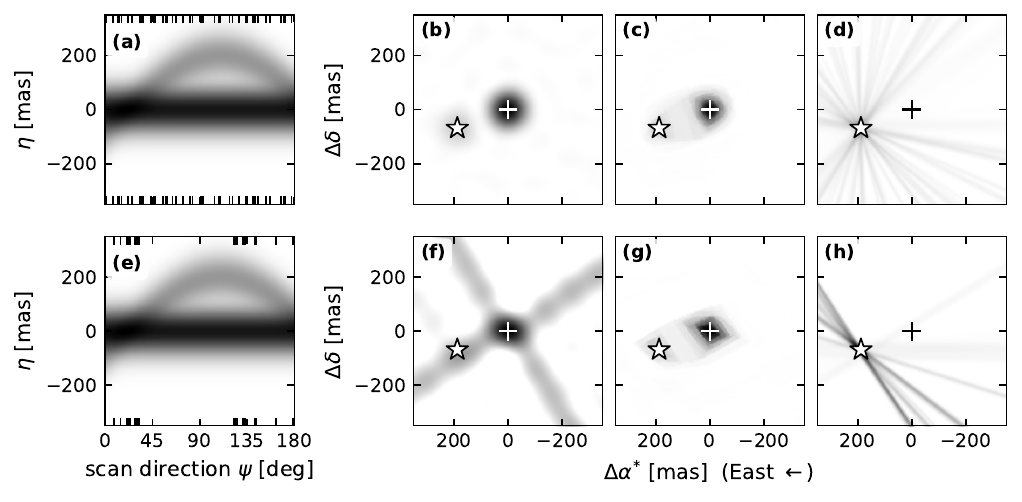}
\caption{Radon inversion, faststack and the triangulation map for the pair of Figure~\ref{fig:scan_to_skyline}, separated by $200$ mas at a position angle of $110^\circ$ with $\mathcal{S}=0.2$ and S/N $100$, on two scan laws. The top row uses the GOST forecast at ecliptic latitude $+60^\circ$, with $95$ transits, and the bottom row the forecast at the Pleiades of Section~\ref{sec:static-example}, with $47$ transits. (a, e) The noise-free profile of the scene against the scan direction over the half-turn, with ticks on the top and bottom edges at the directions that each law samples. (b, f) The ramp-filtered backprojection of the measured windows and (c, g) faststack of the same windows, each linear from zero (white) to its peak. The white plus marks the primary and the star the injected companion. (d, h) The triangulation map of Section~\ref{sec:bp-detect} from the same windows, with the primary fitted and removed in every window and one band per detection. The black plus marks the primary. East is to the left. The companion stands above the spurious level, the largest value of the picture farther than $1.5$ FWHM from both sources, by $8.1$ in the inversion and $21.5$ in faststack on the $+60^\circ$ law, and by $1.1$ and $14.2$ on the Pleiades law, and the peak of the map lies within $5$ mas of the companion on both.}
\label{fig:radon}
\end{figure*}

Faststack \citep{Harrison2011}, the reconstruction that SEAPipe builds \citep{Harrison2023}, is a robust stack of the window samples. It applies no filter and no inversion, so it has no streaks and no negative troughs, but it is not a calibrated image. It reads one sample per window at the projected position of each pixel, so its value changes in steps.

Figure~\ref{fig:radon} applies the inversion and faststack to the pair of Figure~\ref{fig:scan_to_skyline} on two scan laws, and shows the triangulation map of the same windows beside them. At ecliptic latitude $+60^\circ$, no gap between the sampled directions exceeds $10^\circ$, and the inversion returns the blurred scene. The spurious level of a picture is its largest value farther than $1.5$ FWHM from both sources, and there the companion stands $8.1$ times above it. At the Pleiades, the $90^\circ$ gap of Section~\ref{sec:static-example} smears the primary into four arms along the two clusters of directions, and the companion becomes a knot on one of them, at $1.1$ times the spurious level. Faststack shows the companion as a shoulder on the east side of the primary on both laws, at $21.5$ and $14.2$ times the spurious level, because it stacks the samples and never filters them, so a gap in the directions removes coverage but adds no streaks. The map puts the crossing of its bands at the companion on both laws, within $5$ mas of the injected position, because a detection constrains a full line on the sky whatever the gaps between the directions. This inversion is the plain filtered backprojection with one weight per direction. The arms are the streaks of limited-angle tomography, whose geometry is known \citep{FrikelQuinto2013}, and iterative reconstructions suppress them under a prior on the image, total variation for sparse views \citep{SidkyPan2008} or point sources in the CLEAN algorithm of radio interferometry \citep{Hogbom1974}. These methods fill the missing directions from the prior, not from the data. An image of the field from a sky survey that resolves it, from the ground or from space, can supply that prior, and sources it catalogs can be modeled and subtracted from the windows before the inversion. The map needs no prior. The inversion needs the directions, whereas faststack and the map need only the windows.

\begin{figure*}[p]
\centering
\includegraphics[width=0.92\textwidth]{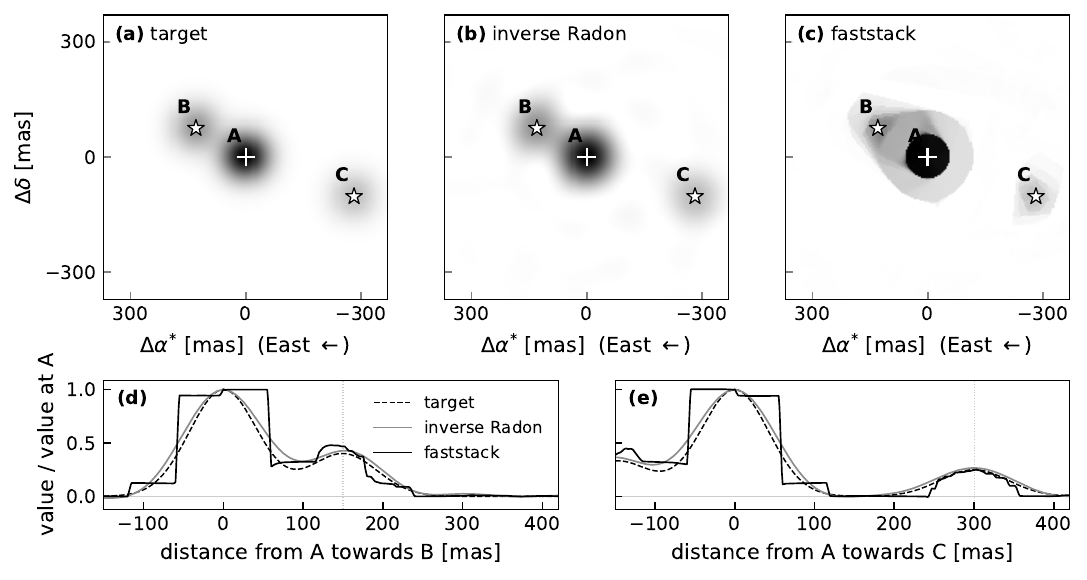}
\caption{Three pictures of one set of windows for a scene of three sources. A is at the origin, B at $150$ mas at a position angle of $60^\circ$ with a flux ratio of $0.40$, and C at $300$ mas at $250^\circ$ with $0.25$, observed at $36$ directions with a pitch of $5^\circ$, one sampling phase and S/N $100$. (a) The target, the scene convolved with the LSF. (b) The inverse Radon transform. (c) Faststack \citep{Harrison2011}. Each picture is divided by its value at A and shown on one linear gray scale from zero (white). (d, e) Cuts from A through B and from A through C, for the target (dashed), the inverse Radon transform (gray) and faststack (black). The pictures give $0.40$, $0.43$ and $0.47$ at B and $0.25$, $0.27$ and $0.24$ at C, and their largest excursions farther than $1.5$ FWHM from the sources are $0.002$, $0.032$ and $0.018$.}
\label{fig:pictures}
\end{figure*}

Figure~\ref{fig:pictures} compares the target, the inversion and faststack for a scene of three sources observed at $36$ evenly spaced directions. With full and even coverage, both image routes return the scene. The inversion returns the target with troughs down to $3\%$ of the value at the primary and a spurious level of $3\%$, and faststack returns the three sources with a spurious level of $2\%$, but with steps and a plateau around the primary. The positions and motions in this paper come from the joint window fit and depend on neither picture.

\end{appendix}

\end{document}